\documentclass[aps,twocolumn,showpacs,superscriptaddress,groupedaddress,floatfix]{revtex4-2}  
\usepackage{graphicx}  
\usepackage{dcolumn}   
\usepackage{bm}        
\usepackage{amssymb}
\usepackage{amsmath}   
\usepackage{hyperref}

\allowdisplaybreaks

\hypersetup{colorlinks,linkcolor={blue},citecolor={blue},urlcolor={red}}
\newcommand{\Mpl}{M_{_{\rm Pl}}}

\def\beq{\begin{equation}}
\def\eeq{\end{equation}}
\def\bea{\begin{eqnarray}}
\def\eea{\end{eqnarray}}

\newcommand{\viz}{\textit{viz.~}}

\begin{document}

\title{Inflation vs. Ekpyrosis --- comparing stability in general non-minimal theory}
\author{Debottam Nandi}
\email{dnandi@physics.du.ac.in}
\author{Manjeet Kaur}
\email{mkaur1@physics.du.ac.in}
\affiliation{Department of Physics and Astrophysics, University of Delhi, Delhi 110007, India}

\begin{abstract}
The scalar field is considered to have dominated the early Universe. One subtle yet crucial requirement of this assumption is that the solution must be highly stable, i.e., indifferent to any initial conditions because there are no favored ones. Inflation, which is now the most successful early Universe paradigm, answers most of the early Universe's problems, including the fact that it is mostly stable. In this article, in addition to the inflationary solution, we systematically investigate every possible early Universe solution in the presence of a barotropic fluid in the general non-minimal (scalar-tensor) theory. In doing so, we rely upon the classical perturbative techniques. We find, to our surprise, that inflation does not always ensure stability in the Einstein frame, although ekpyrosis can. We also discover that, contrary to the inflationary paradigm, ekpyrosis always assures stability in the presence of any fluid with any equation of state in general non-minimal models. We utilize the conformal transformation to map the inflationary theory in the minimal frame to the ekpyrotic theory in the non-minimal frame, and show that the latter is always much more stable than the former, resulting in a much more preferred model that can even be studied in different contexts such as late time cosmology.
\end{abstract}

\maketitle

\section{Introduction}
Stable (attractor) solutions are crucial for solving equation in general relativity because they enable highly nonlinear Einstein equations to asymptotically converge to a particular solution regardless of the initial conditions or the presence of additional matter in the system. For this reason, attractor solutions are consistently favored over non-attractor solutions. These benefits particularly establish themselves in a highly critical position during the early Universe because there are no favorable initial conditions available to us. This means that different initial conditions lead to different solutions in the early Universe, which is undesirable unless the result is an attractor.

At the very early stage of the Universe, it is assumed to be dominated by the scalar field(s) and the most successful theory of it by far is the inflationary paradigm \cite{STAROBINSKY198099, Sato:1981, Guth:1981, LINDE1982389, Albrecht-Steinhardt:1982, Linde:1983gd, Mukhanov:1981xt, HAWKING1982295, STAROBINSKY1982175, Guth:1982, VILENKIN1983527, Bardeen:1983, Starobinsky:1979ty}, in which the Universe expands at an accelerated rate, thus solving early Universe puzzles like the horizon or the flatness problem. The beauty of inflation is, not only it satisfies the observational constraints \cite{Akrami:2018odb,Aghanim:2018eyx}, but also acts as a stable attractor solution \cite{Copeland:1997et, Ng:2001hs}. As a result, perturbations, which are generated quantum mechanically and inherent to the system, do not grow beyond a specific threshold, rendering the system stable and free of different instabilities. Anisotropic stress, which at that high energy is always assumed to be present, is also diminished over this period. Certain concerns, however, like as the trans-Planckian or the initial singularity problem \cite{Martin:2000xs, Borde:1993xh, Borde:2001nh, Lesnefsky:2022fen}, ruling out of inflationary models \cite{Martin:2010hh,Martin:2013tda,Martin:2013nzq,Martin:2014rqa}, etc., plague the inflationary scenario and compel us to look beyond this paradigm.

On the other hand, well-known alternatives to inflation, such as the bouncing paradigm \cite{Novello:2008ra, Cai:2014bea, Battefeld:2014uga, Lilley:2015ksa, Ijjas:2015hcc, Brandenberger:2016vhg}, in which the Universe initially contracts to a minimum and then expands, can also address the problems associated with the early Universe, such as the Horizon and flatness problems without the trans-Planckian problem or the initial singularity problem. However, the greatest challenge lies in constructing a viable bouncing model that satisfies observational restrictions \cite{Cai:2009fn, Gao:2014eaa, Gao:2014hea, Quintin:2015rta, Li:2016xjb, Akama:2019qeh, Kothari:2019yyw}. It also has problems with ghost and gradient instabilities \cite{Kobayashi:2016xpl, Libanov:2016kfc, Ijjas:2016vtq, Banerjee:2018svi, Cai:2016thi, Cai:2017dyi, Kolevatov:2017voe, Mironov:2018oec, Easson:2011zy, Sawicki:2012pz}. However, the issue of stability is of the utmost importance, i.e., in general, bouncing solutions are unstable, and as a result, intrinsic perturbations or anisotropies (Belinsky-Khalatnikov-Lifshitz (BKL) instability) can accumulate and make the system extremely unstable \cite{doi:10.1080/00018737000101171, Karouby:2010wt, Karouby:2011wj, Bhattacharya:2013ut, Cai:2013vm, Ganguly:2021pke}. These issues lead to in search for non-minimal theories, \viz the Horndeski theories or even beyond the Horndeski theories \cite{Horndeski:1974wa, Gleyzes:2014dya, Kobayashi:2019hrl, Cai:2013vm, Kobayashi:2016xpl, Libanov:2016kfc, Ijjas:2016vtq, Banerjee:2018svi, Cai:2016thi, Cai:2017dyi, Kolevatov:2017voe, Mironov:2018oec, Kobayashi:2016xpl, Quintin:2015rta, Li:2016xjb, Cai:2012va, Ilyas:2020qja, Dobre:2017pnt, Zhu:2021whu}.

Nevertheless, there appears to be an exception to this rule: a very slowly contracting bouncing scenario, known as the ekpyrotic bounce \cite{Levy:2015awa}, can avoid this issue even more effectively than the inflationary paradigm. Using non-perturbative simulations, the authors of Ref. \cite{East:2015ggf} demonstrated that the inflationary expansion begins under very precise conditions, casting doubt on the claim that inflation is insensitive to initial conditions. In Ref. \cite{Clough:2016ymm}, the authors showed that under the majority of initial conditions, a small field potential is insufficient to initiate inflation. In fact, authors demonstrated in Ref \cite{Garfinkle:2008ei} that the ekpyrotic contraction is ``super-smooth," i.e., it is resistant to a broad variety of initial conditions and avoids ``Kasner/mixmaster" chaos. A similar assertion on any other primordial scenario could never be substantiated. Also, using classical perturbative analysis, it is possible to demonstrate that inflation is incapable of suppressing cosmological constant-like fluids, but ekpyrosis, on the other hand, can dilute any fluid with any equation of state parameter. In contrast to the inflationary paradigm (mostly slow-roll inflation), however, ekpyrotic bounce cannot produce outcomes that are consistent with observations.

Therefore, the motivation of this article is in uncovering options in quest of a model that may simultaneously suppress the presence of any additional fluid(s) while satisfying observations. Clearly, finding such a model within the framework of minimal Einstein's theory is exceedingly challenging. Once we depart from this, however, and reach the realm of non-minimal theory, it may be possible to find such a model. This is because, as demonstrated in Ref. \cite{Nandi:2018ooh, Nandi:2019xlj}, it is possible to transform any bouncing solution into an attractor solution via non-minimal coupling. Moreover, we showed that these models are also capable of satisfying observational constraints \cite{Nandi:2019xag}. Upon further investigation, we discovered that conformal transformation is the key, since it preserves the invariance of the perturbations but, at the same time, can modify the stability of the system, potentially making it more stable and hence, preferred than the original model. Using these insights, in Ref. \cite{Nandi:2020sif, Nandi:2020szp, Nandi:2022twa}, we constructed bouncing models in the non-minimal frame and have shown that they satisfy the observations and are free of instabilities. However, the question is always finding models with the best apparent stability and observational viability to date.

To find the answers of the above questions, in this work, we extensively study the stability in the context of the early Universe. In this regard, we consider general non-minimally coupled (with scalar field) gravity model in the presence of an additional fluid. Here, for simplicity, we consider the fluid to be barotropic in nature, i.e., the pressure is proportional to the energy density of the system. In doing so, we choose a wide range of equation of state parameter $w_m$, i.e., the ratio of pressure to the energy density from $w_m = -1$ to $w_m = 1$, and study the stability analysis. One could wonder why there is such a large selection of $w_m$. This is because, as far as we are aware, fields in our universe --- whether they be associated with dark energy ($w_m=-1$), dark matter ($w_m=0$), radiation ($w_m=1/3$) or the stiff matter ($w_m=1$) --- contain such equations of state parameters.

Therefore, this work has a dual purpose. The first is to compare the stability of two non-minimal theories that are related by a conformal transformation, and the second is to build up the non-minimal frame in such a way that we can discover a more preferred model in terms of stability while maintaining the observations to be invariant. Conformally connected theories, i.e., two theories associated by the remapping of the metric tensors, are generally assumed to be indistinguishable (equivalent) as long as perturbations remain invariant. In the first half of this paper, we attempt to show a similar relation (equivalence) in the context of stability. In doing so, we first examine the stability in the non-minimal frame without the presence of an additional fluid, and we establish the relationship between the stability criteria in various conformal frames, which has been thoroughly demonstrated in a different way in Ref. \cite{Nandi:2019xlj}. Then, the conformal invariance is deliberately disrupted by introducing the additional fluid, and the dynamical analysis is performed once more. As expected, we discover that the conformal stability criterion is now violated, and we compare the stability in various non-minimal frames once more. We find that, regardless of the presence of non-minimal couplings, the ekpyrotic scenario is always favored because it can dilute any presence of the additional fluid. It can even diminish the effect of the cosmological constant, which may provide a solution to the age-old conundrum of dark energy's fine-tuning. Therefore, an ideal illustration of such a situation would be the following: it is possible to design a non-minimal model that conforms to any viable inflationary theory and change it into an ekpyrotic scenario, making it significantly more desirable than the original inflationary theory. There may be other countless benefits, and it may also be possible to study it in the context of the late universe. However, this is outside the scope of the present project, so we will save it for future initiatives.

The article is written in the following way. In the next section \ref{sec:2-dynamical-eq}, we set up the equations needed for dynamical analysis in the most generalized manner. In Sec. \ref{sec:3-conf-trans}, we introduce the conformal transformation and show how different non-minimal theories are connected in the new set up established for the dynamical analysis. Then, in Sec. \ref{sec:4-without-Baro}, we study the dynamical analysis in different non-minimally coupled models without the presence of the barotropic fluid and we find the equivalence of stability in these theories. In the following section \ref{sec:4-without-Baro}, we again study the dynamical analysis, now in the presence of the barotropic fluid and compare our results in different non-minimally coupled theories. We find the best result lies in ekpyrosis scenario. At the end, we summarize and conclude the final remarks in Sec. \ref{sec:6-conclu}.

In this work, we work with natural units of  $\hbar=c=1$ and we define the Planck mass to be $\Mpl \equiv (8\,\pi\, G)^{-1/2}= 1$. We adopt the metric signature of $(-,+,+,+)$. The partial derivatives are expressed as $\partial$. The overdot $( \dot{\, \, })$  and the subscript $N \equiv \log(a)$ denote derivative with respect to cosmic time $t$ and the e-fold variable $N$, respectively, associated with the Friedmann-Lema$\hat{i}$tre-Robertson-Walker (FLRW) line element, respectively with $a$ being the scale factor and $H$ being used as the Hubble parameter, defined as $\left(\frac{\dot{a}}{a}\right).$ We have normalized all the scalar fields in this work with $\Mpl$.

\section{Dynamical equations in general non-minimal theory}\label{sec:2-dynamical-eq}

Let us now discuss the general equations required to perform such analysis. In this work, we focus on the theories where the action consisting of a scalar field non-minimally coupled to gravity with the presence of an additional barotropic fluid as
\begin{eqnarray}\label{eq:non-minimal-action}
	\mathcal{S} &=& \frac{1}{2} \int {\rm d}^4{\rm \bf x} \sqrt{-g} \left\{ f^2(\phi)\, R - g^{\mu \nu}\, \omega(\phi)\, \partial_\mu \phi \partial_\nu \phi \right.\nonumber\\
	&& \hspace*{75pt}\left. - 2\,V(\phi)\right\} + S_m (g_{\mu \nu}, \Psi_m),
\end{eqnarray}
where, $g_{\mu \nu}$ is the metric tensor, $R$ is the Ricci scalar, $f(\phi)$ is the non-minimal scalar coupling function, $\omega(\phi)$ is the derivative coupling function of $\phi$, $V(\phi)$ is the scalar potential function, and  $S_m (g_{\mu \nu}, \Psi_m)$ is the action associated with the additional barotropic fluid. The corresponding equations of motion can be written as
\begin{eqnarray}
	&& f^2 \left(R_{\mu \nu} -  \frac{1}{2} g_{\mu \nu} R\right) - 2 \nabla_{\mu} f\, \nabla_{\nu} f - 2 f\,  \nabla_{\mu \nu} f  + \nonumber\\
	&&  2 g_{\mu \nu}\,\left( \nabla^\lambda f\, \nabla_\lambda f +  f \,\Box f \right) \  =  \omega(\phi) \left(\nabla_{\mu}\phi \nabla_{\nu}{\phi} - \right.\nonumber\\
	\label{eq:EinsEq}
	&&\hskip60pt\left.\frac{1}{2} g_{\mu \nu} \nabla^\lambda \phi \nabla_\lambda \phi \right) - g_{\mu \nu} V + T^{(m)}_{\mu \nu}, \\
	\label{eq:Scalar Eq}
	&& \Box \phi + \frac{1}{2 \omega(\phi)} \left(\omega_{,\phi}\, \nabla^\lambda\phi \nabla_\lambda\phi - 2 V_{,\phi} + 2 \, f\, f_{,\phi}\, R\right) = 0,\qquad
\end{eqnarray} 
where, $A_{, \phi} \equiv \left(\partial A/\partial\phi\right),$ $\Box \equiv g^{\mu \nu}\nabla_{\mu}\nabla_{\nu}$, $R_{\mu \nu}$ is the Ricci tensor, $R \equiv g^{\mu \nu} R_{\mu \nu}$ is the Ricci scalar and $T^{(m)}_{\mu \nu}$ is the energy-momentum tensor associated with the barotropic fluid satisfying the continuity equation
\begin{eqnarray}\label{eq:continuity}
	\nabla^{\mu} T^{(m)}_{\mu \nu} = 0.
\end{eqnarray}

\noindent Using the Friedmann-Lema\^{\i}tre-Robertson-Walker (FLRW) line element, describing the homogeneous and isotropic Universe
\begin{eqnarray}\label{eq:FRWLineElement}
	{\rm d}s^2 = -{\rm d}t^2 + a^2(t)\,{\rm d}{\bf x}^2 = a^2(\eta)\left(-{\rm d}\eta^2 + {\rm d}{\bf x}^2\right),
\end{eqnarray}
where, $a(t)$ is the scale factor, the above equations take the following form:
\begin{eqnarray}\label{eq:back00}
	&&3 f^{2} \, H^2 = - 6 H\,f\,f_{,\phi} \dot{\phi} +  \left(\frac{1}{2} \omega\, \dot{\phi}^2 + V + \rho_{M}\right),\\
	\label{eq:backij}
	&&2f^2 \dot{H} = -2\, \left(f_{,\phi}^2 + \, f f_{,\phi\phi}\right)\dot{\phi}^2- \, 2f f_{,\phi}\ddot{\phi} +\, 2H\, f f_{,\phi}\dot{\phi} \nonumber\\
	&&\hspace{40pt}- \, (\omega \dot{\phi}^2 +\, (1+w_m)\rho_m)\\
	\label{eq:eqnfield}
	&& \ddot{\phi} + 3 H \,\dot{\phi} + \frac{1}{2 \omega} \left(\omega_{,\phi} \dot{\phi}^2 + 2V_{,\phi} -2  f f_{,\phi} R \right) = 0, \\
	\label{eq:backmatter}
	&& \dot{\rho}_m + 3 H\left(\rho_m + P_m\right) = 0,
\end{eqnarray}

\noindent where, $\rho_m$ and $P_m$ are the energy density and pressure of the additional fluid, i.e.,
$$T^{(m)}{}^{0}_{~0} \equiv -\rho_m, \quad T^{(m)}{}^i_{~j} \equiv P_m\,\delta^{i}_{j},$$ and the equation of state of the barotropic fluid is
$$w_m \equiv \frac{P_m}{\rho_m} = \mbox{Constant.}$$ For example, if the fluid is dust-matter, the corresponding equation of state is $w_m = 0$ and, by using Eq. \eqref{eq:backmatter}, one can obtain the energy density of the additional fluid as $\rho_m \propto a^{-3}.$ Similarly, $w_m = 1/3$ signifies the additional fluid is radiation-like with $\rho_m \propto a^{-4}.$ Most importantly, stiff anisotropic fluid represents $w_m = 1$ and $\rho_m \propto a^{-6}$.

Let us now define the parameters on which we will analyze dynamical properties. These are

\begin{eqnarray}\label{eq:defgammamu}
	\gamma \equiv \frac{V f_{,\phi}}{f V_{,\phi}},\quad \mu \equiv \frac{V\sqrt{\omega}}{f V_{,\phi}}.
\end{eqnarray}
\noindent These are scalar field functions in general. In our case, we treat them as constants because, as we'll see later, they make it easier to find fixed point solutions, which are otherwise quite difficult to obtain. In the case of slow-roll, for example, these functions vary, but the change in the slow-roll limit is insignificant, therefore they can again be considered to be constants.

We can further simplify background equations by defining two dimensionless quantities as:
\begin{eqnarray}\label{eq:xy}
	x \equiv \sqrt{\frac{\omega}{6}}\frac{\dot{\phi}}{H f}, \quad y \equiv \frac{\sqrt{V}}{\sqrt{3}fH}.
\end{eqnarray}
Using the above definitions of $x$ and $y,$ along with $\gamma$ and $\mu,$ the energy equation \eqref{eq:back00} can now be written as
\begin{eqnarray}\label{eq:ham_constraint}
	\Omega_m \equiv \frac{\rho_m}{3  f^2H^2 } = 1 + 2\, \frac{\sqrt{6}}{\mu} \gamma\, x-y^2 -x^2 ,
\end{eqnarray}
where, $\Omega_m$ is the fractional energy density of the additional fluid. One can also express the (first) slow-roll parameter in terms of $x$ and $y$ as
\begin{eqnarray}\label{eq:slow-roll-xy}
	\epsilon \equiv -\frac{\dot{H}}{H^2} &=&\frac{1}{2 \left(6 \gamma ^2+\mu^2\right)}\left(3 \mu ^2 \left(-w_m x^2-(w_m+1) y^2\right.\right.\nonumber\\
	&& \left.\left.+w_m+x^2+1\right)+2 \sqrt{6} \gamma  \mu  (3
	w_m-1) x\right.\nonumber\\
	&&\left.+6 \gamma  \left(2 \gamma  \left(x^2+2\right)-y^2\right)\right),
\end{eqnarray}
as, for constant $\epsilon$, it defines the scale factor solution, i.e., $a(\eta) \propto (-\eta)^{1/(\epsilon - 1)}.$  Finally, the effective equation of state of the Universe in terms of the slow-roll parameter can then be written as
\begin{eqnarray}\label{eq:eeos}
	w_{\rm eff} = -1 + \frac{2}{3}\,\epsilon,
\end{eqnarray}
which essentially signifies how the effective energy density of the Universe depends on the scale factor, i.e.,
\begin{eqnarray}
	\rho_{\rm eff} \propto a^{-3(1 + w_{\rm eff})}.
\end{eqnarray}
As the system contain two scalar degrees of freedom ($\phi$ and the barotropic fluid), the above two dimensionless variables can be used as dynamical quantities, and the equations of motion of $x$ and $y$ can be obtained as
	\begin{eqnarray}\label{eq:eomx}
		\frac{{\rm d} x}{{\rm d}N} &\equiv& \frac{1}{H} \frac{{\rm d} x}{{\rm d}t}= -\frac{1}{2 \left(6 \gamma ^2 \mu +\mu ^3\right)}\left(-3 x^3 \left(4 \gamma ^2 \mu - \mu ^3 \right.\right.\nonumber\\
		&&\left.\left.(w_m-1)\right)+\sqrt{6} \gamma  x^2 \left(24 \gamma
		^2+\mu ^2 (7-9 w_m)\right)\right.\nonumber\\
		&&\left.+6 \gamma  \mu  x \left(6 \gamma  w_m+y^2\right)+3 \mu ^3 x
		\left(w_m \left(y^2-1\right) \right.\right.\nonumber\\
		&& \left.\left.+y^2+1\right)+\sqrt{6} \mu ^2 \left(\gamma  (3 w_m-1)\right.\right.\nonumber\\
		&&\left.\left.+y^2 (1-3
		\gamma  (w_m+1))\right)\right),\\
		\label{eq:eomy}
		\frac{{\rm d} y}{{\rm d}N} &\equiv& \frac{1}{H} \frac{{\rm d} y}{{\rm d}t} 
		= \frac{y}{2 \left(6 \gamma ^2 \mu +\mu ^3\right)} \left(3 x^2 \left(4 \gamma ^2 \mu -\right.\right.\nonumber\\
		&&\left.\left.\mu ^3 (w_m-1)\right)+\sqrt{6} x \left(6 (1-2 \gamma )
		\gamma ^2+\right.\right.\nonumber\\
		&& \left.\left.\mu ^2 (\gamma  (6 w_m-4)+1)\right)-3 \mu ^3 (w_m+1) \left(y^2-1\right)\right.\nonumber\\
		&&\left.-6 \gamma
		\mu  \left(y^2-4 \gamma \right)\right),
	\end{eqnarray}
where, instead of the cosmic time, we have expressed the time variable as $N$, the e-folding number defined as the logarithmic change of scale factor, i.e., $N \equiv \ln{(a)}.$ Notice that, only if $\gamma$ and $\mu$ are (almost) constants can the preceding relations be obtained in terms of $x$ and $y$, resulting in fixed point (analytically solvable) solutions. One can debate the legitimacy of such a choice. However, one might counter the notion that because this is always achievable in a short range of time, they can be deemed constant.

Since we now have the equations in terms of the dimensionless variables $x$ and $y$, we may solve the system, which will be discussed in the subsequent sections. However, before we do that, we will explore the conformal transformation and the relationship between two conformally connected frames in the following part, which can assist us in establishing the relationships of the scale factors between these two frames.

\section{Conformal transformation}\label{sec:3-conf-trans}

In this section, given a theory in expression \eqref{eq:non-minimal-action}, we apply the conformal transformation on that action, with the goal being to discover connections between various non-minimal frames, as was previously described. To ensure that conformal invariance breaks down between the two models and the physics differ in them, we just transform the scalar part of the action, leaving the additional fluid unaffected. In the conformal transformation, the metric field as well as the scale factor are redefined as
\begin{eqnarray}\label{eq:ConfTGen}
	\tilde{g}_{\mu \nu} = \Omega^2(\phi)\,g_{\mu \nu} \quad \Rightarrow \quad \tilde{a}(\tilde{t}) = \Omega(\phi)\,a(t).
\end{eqnarray}
$\tilde{g}_{\mu \nu}$ and $\tilde{a}(\tilde{t})$ are the new conformally transformed metric and the scale factor, respectively, describing a new Universe, whereas, $g_{\mu \nu}$ and $a(t)$ are the old metric and the scale factor, respectively. Also, the cosmic time is replaced by the new cosmic time parameter $\tilde{t}$, where they are related by ${\rm d} \tilde{t}  = \Omega\,{\rm d}t.$ Under such transformation, the non-minimal theory described in \eqref{eq:non-minimal-action} transforms into a new action 

\begin{eqnarray}\label{eq:non-min-trans}
	\tilde{\mathcal{S}} &=& \frac{1}{2} \int {\rm d}^4{\rm \bf x} \sqrt{-\tilde{g}} \left[ \tilde{f}^2(\phi)\,\tilde{R} - \tilde{\omega}(\phi) \,\tilde{g}^{\mu \nu} \partial_\mu \phi \partial_\nu \phi \right.\nonumber\\
	&& \qquad \qquad \qquad\qquad \qquad\left.   - 2\, \tilde{V}(\phi)\right] + S_m.
\end{eqnarray} 
Again, please note that $S_m$ represents the identical action for the barotropic fluid defined in the original action in  \eqref{eq:non-minimal-action} as we are conformally transforming only the scalar sector of the action. The other functions, $\tilde{f}(\phi),\, \tilde{\omega}(\phi)$ and the potential $\tilde{V}(\phi)$  depend on the coupling functions $f(\phi),~ \omega(\phi)$ and the potential $V(\phi)$ as
\begin{eqnarray}\label{eq:conff}
	\tilde{f}(\phi) &=& \frac{f}{\Omega},\\
	\label{eq:confomega}
	\tilde{\omega}(\phi) &=& \left(\frac{\omega}{\Omega^2}+6\frac{f^2}{\Omega^2}\left(2\frac{f_{, \phi}}{f}-\frac{\Omega,_{ \phi}}{\Omega}\right)\frac{\Omega,_{ \phi}}{\Omega}\right),\\
	\label{eq:confpot}
	\tilde{V}(\phi) &=& \frac{V}{\Omega^4}.
\end{eqnarray}
The corresponding equations of motion of this theory is identical to Eqs. \eqref{eq:EinsEq} and \eqref{eq:Scalar Eq} with the old functions replaced by the functions defined in the new conformally transformed theory with new time variable $\tilde{t}.$ Note that, in the new transformed theory, the new $\tilde{\gamma}$ and $\tilde{\mu}$ are defined, again, as

\begin{eqnarray}\label{eq:confgammamu}
	\tilde{\gamma} \equiv \frac{\tilde{V} \tilde{f}_{,\phi}}{\tilde{f} \tilde{V}_{,\phi}},\quad \tilde{\mu} \equiv \frac{\tilde{V}\sqrt{\tilde{\omega}}}{\tilde{f} \tilde{V}_{,\phi}}.
\end{eqnarray}
As a result, the dynamical equations given in Eqs. \eqref{eq:eomx} and \eqref{eq:eomy} remain identical in the new frame, with all quantities replaced by newly specified functions in the new frame, including $\tilde{\gamma}$ and $\tilde{\mu}.$ The above relation shows the link between $\{\gamma, \mu\}$ and $\{\tilde{\gamma}, \tilde{\mu}\}$, and because dynamics rely on them, we can now build a direct relationship between two frames, which will be realized as we progress.

Since the set up is now developed, we now can dynamically analyze the system. First, in the next section, we will consider the action without the additional fluid, and later, we will focus on the effect of the additional fluid causing the breaking of conformal invariance to understand the complete picture.


\section{With the absence of barotropic fluid}\label{sec:4-without-Baro}

In the absence of the additional fluid, $\rho_m = 0$, which leads to
\begin{eqnarray}
	\Omega_m = 0, \quad \Rightarrow y^2 = 1 + 2\, \frac{\sqrt{6}}{\mu} \gamma\, x - x^2.
\end{eqnarray}
It implies that the two variables depend on each other, which essentially removes one degree of freedom, which is obvious as there is no additional fluid. Considering $x$ to be the dynamical variable, we can now obtain the evolution equation of it by replacing $y$ in terms of $x$ in Eq. \eqref{eq:eomx} and it becomes
	\begin{eqnarray}\label{eq:eomdelx}
		\frac{{\rm d}x}{{\rm d} N} &=& \frac{1}{2 \left(6 \gamma ^2 \mu +\mu ^3\right)}\left(\sqrt{6} (4 \gamma -1) \mu ^2+6 \mu  x^3 \left(2 \gamma ^2 \right.\right.\nonumber\\
		&&\left.\left.+\gamma+\mu ^2\right)+\sqrt{6} x^2
			\left(\mu ^2-4 \gamma  \left(6 \gamma ^2+3 \gamma +4 \mu ^2\right)\right)\right.\nonumber\\
			&&\left.-6 \mu  x \left(-6 \gamma
			^2+3 \gamma +\mu ^2\right)\right).
	\end{eqnarray}

\noindent The fixed points can be found by equating the velocity ${\rm d}x/{\rm d}N$ to zero, which, here, leads to three critical points:

\begin{eqnarray}\label{eq:fp1wf}
	x_1^* &=& \frac{(4 \gamma -1) \mu }{\sqrt{6} \left(2 \gamma ^2+\gamma +\mu ^2\right)}, \\
	\label{eq:fp2wf}
	x_2^* &=& \frac{\sqrt{6} \gamma -\sqrt{6 \gamma ^2+\mu ^2}}{\mu }, \\
	\label{eq:fp3wf}
	x_3^* &=& \frac{\sqrt{6 \gamma ^2+\mu ^2}+\sqrt{6} \gamma }{\mu },
\end{eqnarray}
with the corresponding slow-roll parameters
\begin{eqnarray}\label{eq:e1wf}
	\epsilon_1 &=& \frac{8 \gamma ^2-6 \gamma +1}{2 \left(2 \gamma ^2+\gamma +\mu ^2\right)} \\
	\label{eq:e2wf}
	\epsilon_2 &=& 3 + \frac{2 \gamma  \left(6
		\gamma -\sqrt{6} \sqrt{6 \gamma ^2+\mu ^2}\right)}{\mu ^2} \\
	\label{eq:e3wf}
	\epsilon_3 &=& 3 + \frac{2 \gamma  \left(\sqrt{6} \sqrt{6
			\gamma ^2+\mu ^2}+6 \gamma \right)}{\mu ^2}
\end{eqnarray}
\noindent Notice that, the second and third fixed point correspond to kinetic energy dominated solutions as these lead to
\begin{eqnarray}
	y_2^* = y_3^* = 0, 
\end{eqnarray}
i.e., the potential energy is zero. Instead, we're interested in solutions that involve potential energy, therefore we focus on the first fixed point \eqref{eq:fp1wf}, which is critical to the dynamics of the early Universe. Please bear in mind that, even if an additional fluid is present, we will always evaluate only this fixed point because it reflects our desired scalar field dominated solution in the early Universe. 

Let us now find out the solution of around the fixed point. To obtain this, we assume linear perturbation theory, i.e., the deviation from the fixed point solution is small: $x = x_* + \delta x, ~\delta x \ll x_*.$ One can then immediately find the equation of motion of $\delta x$ to be

\begin{eqnarray}\label{eq:linwf}
	\frac{{\rm d}\delta x}{{\rm d} N} = \frac{\partial A(x)}{\partial x}\Big{|}_*\,\delta x,
\end{eqnarray}
where, $A(x)$ is the right hand side of Eq. \eqref{eq:eomdelx} and $|_*$ represents the value evaluated at the corresponding fixed point. Keep in mind that here we ignore the higher-order terms of $\delta x$ because we are treating $\delta x$ as a perturbed quantity. In a nutshell, we use the classical perturbation analysis to evaluate the dynamics of the deviation. Following that, it is simple to assess the solution of the above equation as
\begin{eqnarray}
	\delta x = \delta x_0 \exp\left(\lambda N\right), \quad \lambda \equiv \frac{\partial A(x)}{\partial x}\Big{|}_*,
\end{eqnarray}
and for the three fixed points, the eigenvalues, referred to as the Lyapunov exponents, take the following forms:
\begin{eqnarray}\label{eq:l1wf}
	\lambda_1 &=& -3 +\frac{1-2 \gamma  (4 \gamma +1)}{2 \left(2 \gamma ^2+\gamma +\mu ^2\right)},\\
	\label{eq:l2wf}
	\lambda_{2} &=& 6-\frac{6 (2 \gamma +1)}{\sqrt{6} \sqrt{6 \gamma ^2+\mu ^2}+6 \gamma }, \\
	\label{eq:l3wf}
	\lambda_{3} &=& 6 +\frac{6 (2 \gamma +1)}{\sqrt{6} \sqrt{6 \gamma ^2+\mu ^2}-6 \gamma }.
\end{eqnarray}
Given initial conditions close to the fixed point (i.e., choosing $\delta x_0 \ll x_*$), the appropriate value of $\lambda$ determines whether the initial deviation (from the fixed point) decays over time. If initial conditions are slightly deviated from the fixed point solution, $\lambda$ influences whether the solution grows and moves away from the fixed point solution, known as a repeller solution, or decays and eventually merges with the fixed point solution, known as an attractor solution. Thus, for an expanding universe, $\lambda > 0$ indicates that a fixed point is unstable, whereas $\lambda < 0$ indicates that the point is stable. In the next sections, we will examine the relationship between the solution in the minimal Einstein frame and the non-minimally coupled frame.

\subsection{Minimal Einstein frame}

Let us now concentrate on Einstein's minimal frame. As discussed before, we are now considering only the first fixed point \eqref{eq:fp1wf}. In this instance, the coupling function $f(\phi)$ is constant, and hence, $\gamma = 0$. Consequently, Eqs. \eqref{eq:fp1wf} and \eqref{eq:e1wf} become

\begin{eqnarray}\label{eq:minfp1}
	x_1^* = -\frac{1}{\sqrt{6}\mu}, \quad \epsilon_1 = \frac{1}{2 \mu^2},
\end{eqnarray}

\noindent and the corresponding eigenvalue in Eq. \eqref{eq:l1wf} take the form

\begin{eqnarray}
	\lambda_{1} = -3 + \frac{1}{2 \mu^2}.
\end{eqnarray}
Using Eq. \eqref{eq:minfp1}, it is convenient to express the above relation in terms of the slow-roll parameter as
\begin{eqnarray}\label{eq:minL1wf}
	\lambda_{1} = \epsilon_1 - 3 \quad \Rightarrow \lambda_{1} = \frac{1 - 2 \alpha}{\alpha},
\end{eqnarray}
where, $\alpha \equiv 1/(\epsilon_1-1)$ corresponds to the scale factor solution: $a(\eta) \propto (-\eta)^\alpha$ for the first fixed point. Therefore, if the Universe is expanding, $\lambda < 0$ leads to the stable solution which in turn provide the bound on $\alpha$ as

\begin{eqnarray}
	\alpha < 0, \quad\text{or,}\quad \alpha > \frac{1}{2}.
\end{eqnarray}
Adding the fact that, the solution needs to solve the horizon problem, i.e.,  $\eta$ is negative and $|\eta| \rightarrow 0$ (required solution for the early Universe), it leads to one condition

\begin{eqnarray}
	\alpha < 0.
\end{eqnarray}
As $\alpha$ is negative and approaches zero, it is evident that the solution leads to accelerated expansion, and $ \alpha = -1$ is the well-known de-Sitter expansion, which, when expressed in cosmic time $t$, is purely exponential. Therefore, in the minimal Einstein theory, accelerated expansion behaves as an attractor. In contrast, when contraction occurs, the condition becomes

\begin{eqnarray}
	0 < \alpha < \frac{1}{2},
\end{eqnarray}
which is known as the ekpyrotic solution \cite{Levy:2015awa} --- the explanation for the sole stable bouncing (contracting) solution. 
\subsection{Non-minimal frame}

Now that the minimum condition has been examined, we may shift our attention to conformal transformations of action, with the goal of modifying the scale factor to behave as $\tilde{a}(\eta) \propto (-\eta)^\beta$ in the transformed frame. It's important to keep in mind that the conformal time coordinate, $\eta$, remains invariant by conformal transformation, in contrast to the cosmic time $t$. This new solution for the scale factor can be achieved in the following way.

In the minimal frame, with the help of the fixed point solution in Eq.  \eqref{eq:minfp1} as well as Eq. \eqref{eq:defgammamu}, the conformal time can be written in terms $\phi$ as

\begin{eqnarray}\label{eq:minfp1conftime}
	(-\eta)  \propto V(\phi)^{-\frac{1}{2 (1 + \alpha)}}.
\end{eqnarray}
Then according to Eq. \eqref{eq:ConfTGen}, the coupling function takes the form:

\begin{eqnarray}
	\Omega(\phi) = V(\phi)^{\frac{\alpha - \beta}{2(1 + \alpha)}},
\end{eqnarray}
and, using Eqs. \eqref{eq:conff}, \eqref{eq:confomega} and \eqref{eq:confpot}, we get the following functions in terms of  the scalar field $\phi$ as 
\begin{eqnarray}\label{eq:tilf}
	\tilde{f}(\phi) &=& V(\phi )^{\frac{\beta -\alpha }{2 (\alpha +1)}}, \\
	\label{eq:tilw}
	\tilde{\omega}(\phi) &=& \frac{\left(2 (\alpha +1)^2 \mu ^2-3 (\alpha -\beta )^2\right) V'^2\, V^{\frac{-3 \alpha
				+\beta -2}{\alpha +1}}}{2 (\alpha +1)^2},\qquad \\
			\label{eq:tilV}
	\tilde{V}(\phi) &=& V(\phi )^{\frac{-\alpha +2 \beta +1}{\alpha +1}}.
\end{eqnarray}
These functions define the new action in \eqref{eq:non-min-trans} that leads to the new scale factor solution. Using the above definitions, along with Eq. \eqref{eq:confgammamu}, we get the transformed $\gamma$ and $\mu$, i.e., $\tilde{\gamma}$ and $\tilde{\mu}$ in terms of $\alpha$ and $\beta$ as,

\begin{eqnarray}\label{eq:gammaconf}
	\tilde{\gamma} &=& \frac{\beta -\alpha }{2 (-\alpha +2 \beta +1)}, \\
	\label{eq:muconf}
	\tilde{\mu} &=& \frac{\sqrt{-2 \alpha ^2+6 \alpha  \beta +\alpha -3 \beta ^2}}{\sqrt{2} (-\alpha +2 \beta +1)}.
\end{eqnarray}
As was previously stated, these relations provide the relationship between the new non-minimal frame and the previous minimal frame. $\alpha$ indicates the scale factor solution in the minimal frame, whereas $\beta$ specifies the new scale factor. Together, they constitute the non-minimal frame \eqref{eq:non-min-trans} expressed within the functions in Eqs. \eqref{eq:tilf}, \eqref{eq:tilw}, and \eqref{eq:tilV}.

As the non-minimal theory is now established, using the expression \eqref{eq:l1wf}, the Lyapunov exponent of the first fixed point can now again be evaluated as
\begin{eqnarray}
	\tilde{\lambda}_1 = \frac{1-2 \alpha }{\beta},
\end{eqnarray}
which seems to be different from the minimal counterpart represented in Eq. \eqref{eq:minL1wf}. The connection between these two distinct eigenvalues will be discussed in the following section.

\subsection{Equivalence of stability}
As shown above, the two theories lead to two different Lyapunov exponent solutions, implying that the decay rate of the deviations is different in these two theories, i.e.,

\begin{eqnarray}
	\delta x \propto e^{\lambda_1 N}, \quad \delta \tilde{x} \propto e^{\tilde{\lambda}_1 \tilde{N}},
\end{eqnarray}
where $\delta \tilde{x}$ signifies the deviation in the conformally modified theory described in \eqref{eq:non-min-trans}. However, it is clear that the e-folding number in two separate frames differ and is related as 

\begin{eqnarray}
	\tilde{N} = \frac{\beta}{\alpha} N
\end{eqnarray}
up to a constant. This quickly results in

\begin{eqnarray}
	\lambda_1 N  = \tilde{\lambda}_1 \tilde{N}, \quad \Rightarrow \delta \tilde{x} \propto \delta x.
\end{eqnarray}
meaning that variations in both frames have the same effect. This is even clearer if we express the time in conformal coordinates $\eta$. In this instance

\begin{eqnarray}
	\delta x \propto \delta \tilde{x} \propto (-\eta)^{(1 - 2 \alpha)}.
\end{eqnarray}
It shows us that, even if the deviations evolve differently in any other coordinate time (cosmic time $t$ or the e-folding number $N$), they behave similarly in conformal time $\eta$, and therefore the equivalence. It in no way implies that the deviations in different frames behave identically, as time runs independently in each frame. What this tells us is that if the conformal evolution of the two theories is equal, then the deviations likewise behave identically. The anticipated outcome is demonstrated in Ref. \cite{Nandi:2019xlj}. Indeed, it has also been shown that the equivalence can be extended to any connected conformal theory as 

\begin{eqnarray}
	\lambda \,{\rm d} N = \tilde{\lambda} \,{\rm d} \tilde{N}.
\end{eqnarray}

\noindent This is the first outcome of our efforts.

\section{With the addition of barotropic fluid}\label{sec:5-with-baro}
Now, as previously indicated, consider the model in the presence of a barotropic fluid. Because fluid is present, unlike in the previous situation, the degrees of freedom are now two, which are represented by $x$ and $y$. Furthermore, only the scalar component is conformally connected, but the barotropic fluid is unaffected by the conformal transformation, implying that the two theories are not conformally associated, as mentioned in Sec. \ref{sec:3-conf-trans}. Fixed points can then be obtained in this scenario by explicitly setting the velocity of these two variables, i.e., ${\rm d}x/{\rm d}N = {\rm d}y/{\rm d}N = 0$, resulting in seven fixed points \cite{Copeland:1997et, Copeland:2006wr, Nandi:2018ooh, Nandi:2019xlj, Nandi:2020sif, Nandi:2020szp, Nandi:2022twa}:
\begin{eqnarray}
		\label{eq:fp1}
		1.~&&x^*_1 = \frac{(-1+4\gamma)\mu}{\sqrt{6}(\gamma+\, 2\gamma^2 +\, \mu^2)}, \nonumber\\
		&&  y^*_1 = \frac{\sqrt{\begin{aligned}
					&48\gamma^3+120\gamma^4+8\gamma\mu^2+\mu^2(-1+6\mu^2)+\nonumber\\
					&\qquad\qquad\qquad\qquad\qquad\qquad
					\gamma^2(-6+56\mu^2)
				\end{aligned}}}{ \sqrt{6}(\gamma+\, 2\gamma^2 +\, \mu^2)},\nonumber\\
			&& \\
		\label{eq:fp2}
		2.~&& x^*_2 = \frac{(-1+4\gamma)\mu}{\sqrt{6}(\gamma+\, 2\gamma^2 +\, \mu^2)}, \nonumber\\
		&& y^*_2 = -\frac{\sqrt{\begin{aligned}
					&48\gamma^3+120\gamma^4+8\gamma\mu^2+\mu^2(-1+6\mu^2)+\nonumber\\
					&\qquad\qquad\qquad\qquad\qquad\quad\gamma^2(-6+56\mu^2)
				\end{aligned}}}{ \sqrt{6}(\gamma+\, 2\gamma^2 +\, \mu^2)},\nonumber\\
			&&\\
		\label{eq:fp3}
		3. ~&&x^*_3 =\frac{\sqrt{6}\gamma -\, \sqrt{6\gamma^2 +\, \mu^2}}{\mu}, \quad y^*_3 = 0, \\
		\label{eq:fp4}
		4. ~&& x^*_4 = \frac{\sqrt{6}\gamma +\, \sqrt{6\gamma^2 +\, \mu^2}}{\mu}, \quad y^*_4 = 0, \\
		\label{eq:fp5}
		5.~ &&x^*_5 =-\sqrt{\frac{3}{2}} (1 + w_m) \mu,\nonumber\\&& y^*_5 =\frac{\sqrt{\begin{aligned}
					&\gamma(2 -\, 6w_m) +\, 12(1 +\, w_m)\gamma^2-\nonumber\\
					&\qquad\qquad\qquad\qquad\, 3 (-1+w_m^2) \mu^2  
				\end{aligned}}}{\sqrt{2}},\nonumber\\
			&&\\
		\label{eq:fp6}
		6.~ &&x^*_6 =-\sqrt{\frac{3}{2}} (1 + w_m) \mu,\nonumber\\&& y^*_6 =-\frac{\sqrt{\begin{aligned}
					&\gamma(2 -\, 6w_m) +\, 12(1 +\, w_m)\gamma^2-\nonumber\\
					&\qquad\qquad\qquad\qquad\, 3 (-1+w_m^2) \mu^2
				\end{aligned}  }}{\sqrt{2}},\nonumber\\
			&&\\
		\label{eq:fp7}
		7.~ &&x^*_7 = \sqrt{\frac{2}{3}}\frac{(1 - \, 3\,  w_m)\gamma\mu}{(4\gamma^2 + \, (1 - \, w_m)\mu^2)}, \quad y^*_7 = 0.
\end{eqnarray}

It is worth noting that $H$ appears in the denominator of the equation $y$ in Eq. \eqref{eq:xy}. As a result, while a positive sign of $y$ indicates that the Universe is expanding, a negative sign indicates that the Universe is contracting and thus the fixed points in the above expressions appear in pairs (e.g., first and second fixed point). Using Eq. \eqref{eq:slow-roll-xy}, one can quickly calculate the slow-roll parameter for each of these fixed positions, which are listed below:
\begin{eqnarray}\label{eq:epfp1}
	\epsilon_1 &=& \frac{(1-6\gamma+8\gamma^2)}{2(\gamma+2\gamma^2+\mu^2)}\\
	\label{eq:epfp2}
	\epsilon_2 &=& \frac{(1-6\gamma+8\gamma^2)}{2(\gamma+2\gamma^2+\mu^2)}\\
	\label{eq:epfp3}
	\epsilon_3 &=& 3+\frac{2\gamma(6\gamma-
		\sqrt{6}\sqrt{6\gamma^2+\mu^2}}{\mu^2}\\
	\label{eq:epfp4}
	\epsilon_4 &=& 3+\frac{2\gamma(6\gamma+
		\sqrt{6}\sqrt{6\gamma^2+\mu^2}}{\mu^2}\\
	\label{eq:epfp5}
	\epsilon_5 &=&-\frac{3}{2}(1+w_M)(-1+2\gamma)\\
	\label{eq:epfp6}
	\epsilon_6 &=&-\frac{3}{2}(1+w_M)(-1+2\gamma)\\
	\label{eq:epfp7}
	\epsilon_7 &=&\frac{3(-1+w_M)\mu^2-16\gamma^2}{2(-1+w_M)\mu^2-8\gamma^2}.
\end{eqnarray}
Among them, only the first, second, third, and fourth solutions, i.e., Eqs. \eqref{eq:fp1}, \eqref{eq:fp2}, \eqref{eq:fp3}, and \eqref{eq:fp4} are scalar field dominated solutions as the fractional energy density in Eq. \eqref{eq:ham_constraint} vanishes for each of these solutions, i.e.,
$$\Omega_{m1} = 0,\quad \Omega_{m 2} = 0, \quad\Omega_{m3}  = 0,\quad \Omega_{m4}=0.$$
The remaining solutions, Eqs. \eqref{eq:fp5}, \eqref{eq:fp6} and \eqref{eq:fp7} --- refer to the mixed state solutions, in which both the scalar field and the barotropic fluid's energy densities are still non-zero:
\begin{eqnarray}
	&&\Omega_{m5} = -6 \gamma^2 (w_m+1)-\gamma  (3 w_m+7)-\nonumber\\
	&&\qquad\qquad \qquad \qquad\qquad\qquad 3 \mu ^2 (w_m+1)+1, \\
	&&\Omega_{m6} = -6 \gamma^2 (w_m+1)-\gamma  (3 w_m+7)-\nonumber\\
	&&\qquad\qquad\qquad\qquad\qquad\qquad3 \mu ^2 (w_m+1)+1,\\
	&& \Omega_{m7} = \frac{\left(6 \gamma ^2+\mu ^2\right) \left(8 \gamma ^2 (2-3 w_m)+3 \mu ^2
		(w_m-1)^2\right)}{3 \left(\mu ^2 (w_m-1)-4 \gamma ^2\right)^2}.\nonumber\\
	&&
\end{eqnarray}
Comparing the above fixed points with the results from the previous section reveals that $\{x_1^*, y_1^*\}$ (Eq. \eqref{eq:fp1}) and $\{x_2^*, y_2^*\}$ (Eq.\eqref{eq:fp2}) are the desired fixed points. Nonetheless, it can be seen that both fixed points are similar (for instance, compare the slow-roll parameters $\epsilon_1$ in Eq. \eqref{eq:epfp1} and $\epsilon_2$ Eq. \eqref{eq:epfp2}), with the exception that the signs of $y$ are opposite, signifying that they represent identical solutions, one with forward-time and the other with backward-time. Now, in order to analyze the stability of these fixed points as we did in Eq. \eqref{eq:linwf}, we must linearize Eqs. \eqref{eq:eomx} and \eqref{eq:eomy} as follows:

\begin{eqnarray}\label{eq:LinearizedEq}
	\left(\begin{aligned}
		& \frac{{\rm d} \delta x}{{\rm d}N}\\
		& \frac{{\rm d} \delta y}{{\rm d}N}
	\end{aligned}\right) = \left(\begin{aligned}
		\frac{\partial A(x, y)}{\partial x}\Big|_* && \frac{\partial A(x, y)}{\partial y}\Big|_* \\
		\frac{\partial B(x, y)}{\partial x}\Big|_* && \frac{\partial B(x, y)}{\partial y}\Big|_*
	\end{aligned}\right)\left(\begin{aligned}
		\delta x &\\\delta y &
	\end{aligned}\right),
\end{eqnarray}
where, $A(x, y)$ and $B(x, y)$ represent the right-hand sides of Eqs. (\ref{eq:eomx}) and (\ref{eq:eomy}), respectively, and $|_*$ represents the value at the fixed point. $\delta x$ and $\delta y$ represent the deviations of $x$ and $y$ from their fixed points $(x_*, y_*)$, respectively. The above-mentioned square matrix must be diagonalized to determine the eigenvalues and eigenvectors, which will help us find the solutions of $\delta x$ and $\delta y$, as the solutions can then be expressed as

\begin{eqnarray}\label{eq:gen-dev-exp}
	&&\delta x = C_{11}\, e^{\lambda_1\, N} + C_{12}\, e^{\lambda_2\,N}, \nonumber \\
	&&\delta y = C_{21}\, e^{\lambda_1\, N} + C_{22}\, e^{\lambda_2\,N}.
\end{eqnarray}
The eigenvalues of the matrix are $\lambda_1$ and $\lambda_2$, known as the Lyapunov exponents, as before, and $C$'s are related to the eigenvectors as well as the initial conditions. In this scenario, if both the eigenvalues are negative (positive) during expansion (contraction), the deviations decay, implying that the fixed point is stable, or an attractor. If, on the other hand, one or both exponents become positive (negative), i.e., the solution is a non-attractor, then $\delta x$ and $\delta y$ grow with time, and the entire solution rapidly moves away from the desired fixed point, and the system may become highly unstable. 

Using the above method, one can evaluate the eigenvalues for all fixed points; however, because we are only interested in the first (and the second) fixed point $\{x_1^*, y_1^*\}$ given in Eqs. \eqref{eq:fp1}, these exponents have the following form:

\begin{eqnarray}\label{eq:LEfp1}
	&&\lambda_1 = -3 - 3 w_m +\frac{1- 4 \gamma}{(\gamma +\, 2\gamma^2+\, \mu^2)},\nonumber\\
	&&\lambda_2 = - 3 +  \frac{(1 - 4 \gamma)(1 + 2 \gamma)}{2(\gamma +\, 2\gamma^2+\, \mu^2)}.
\end{eqnarray}
Now that we have calculated the eigenvalues for the desired fixed point in the general non-minimal frame, we will assess and compare in two separate scenarios, namely the minimal and non-minimal frames, in the following section.

\subsection{Minimal Einstein frame}
In this instance, $\gamma = 0$; hence, first fixed point in \eqref{eq:fp1} (or the second fixed point in \eqref{eq:fp2}) and the slow-roll parameter in Eq. \eqref{eq:epfp1} are expressed as follows:

\begin{eqnarray}
	x_1^* = -\frac{1}{\sqrt{6}\mu}, \quad y_1* = \pm\frac{\sqrt{6 \mu ^2-1}}{\sqrt{6} \mu},\quad \epsilon_1 = \frac{1}{2 \mu^2}.
\end{eqnarray}
The associated Lyapunov exponents in Eq. \eqref{eq:LEfp1} can therefore be derived as

\begin{eqnarray}\label{eq:eigenmin}
	\lambda_{1} = \frac{2 - \alpha (1 + 3 w_m)}{\alpha },\quad \lambda_{2}= \frac{1 - 2\alpha}{\alpha},
\end{eqnarray}
where, $\alpha \equiv 1/(\epsilon_1 - 1),$ yields the scale factor solution, $a(\eta)\propto (-\eta)^\alpha$, as described in the preceding section. Notice that $\lambda_{2}$ is similar to that in the minimal scenario, i.e., Eq. \eqref{eq:l1wf}, without the additional fluid. In contrast, due to the barotropic fluid, we now have a new eigenvalue that likewise depends on its equation of state parameter $w_m$. Immediately, we now have the following stability conditions:
\begin{eqnarray}\label{eq:cond_stab_min_ex}
	&&\text{For expansion:} \begin{cases}
		&w_m < -\frac{1}{3},~\text{and}~ \alpha >\frac{2}{1 + 3 w_m},\\
		&\text{or,}\\
		&w_m \geq -\frac{1}{3},~\text{and}~ \alpha < 0. 
	\end{cases} \nonumber\\
&&\\
&&\text{For contraction:}
	\begin{cases}
		& w_m \leq 1,~ \text{and}\quad \alpha < \frac{1}{2},\\
		&\text{or,}\\
		& w_m > 1, ~\text{and}\quad \alpha < \frac{2}{1 + 3 w_m}.
	\end{cases} \nonumber\\
&&
\end{eqnarray}
Please keep in mind that while evaluating scalar perturbations, we have additional constraint on $\alpha$: $-1 < \alpha < 0$ is forbidden because it leads to ghost instability. However, because $\alpha \leq -1$ indicates that the Universe is in an accelerated (inflationary) expansion, we find that all inflationary solutions are guaranteed to be stable only if $w_m > -1/3$, whereas stability is conditional for $w_m \leq -1/3$, as the lower bound of $\alpha$ is: $\alpha_{\rm min} \equiv 2/(3w_m +1).$ It follows instantly that if the barotropic fluid is de-Sitter, as with $w_m = -1$, $\alpha_{\rm min}$ becomes $-1.$ In the most slow-roll inflationary scenario, at or near the pivot scale, the effective value of $\alpha$ is extremely close to $-1$, i.e., $\alpha \lesssim -1.$ As a classic example, consider chaotic inflation, where $\epsilon_1 \simeq 0.01$ and, consequently, $\alpha \simeq -1.01$ at the pivot scale. In that scenario, as $\alpha < \alpha_{\rm min},$ the slow-roll solution is not a stable solution. When we consider $w_m \leq -1$, the severity of the situation grows.

The following result is not surprising given that, during conventional slow-roll inflation (at or near the pivot scale), the energy density $\rho_{\phi} \sim 10^{-9} \Mpl^4$ steadily decays to one or two orders below that level, at the minimum of the potential, resulting in a (p)reheating scenario (which in the current context, we are not studying). In the presence of a de-Sitter type fluid, however, the energy density of the fluid $\rho_m$ remains constant, and so $\rho_m/\rho_{\phi}$ increases by an order or two during inflation, explaining the instability of the fixed point. As a result, it instantly suggests that, unless $\rho_m \ll \rho_\phi,$ a de-Sitter like fluid with $\rho_m \sim \rho_\phi$ can induce system instability. However, it should be emphasized that any natural energy scale below the Planck level, such as electroweak energy, is considerably below the Hubble scale, and so the $\rho_m \ll \rho_\phi$ condition is satisfied, which does not affect the system.

In contrast, in the case of contraction, $\alpha < 1/2$ (the ekpyrotic solution) always results in stability, regardless of the equation of state of the additional fluid. In reality, as Eq. \eqref{eq:eigenmin} shows, $\alpha \rightarrow 0^+$ renders the system unaffected by $w_m$, regardless of the value as $w_m$ is multiplied by $\alpha$. This conclusion is crucial because it suggests that ekpyrosis can dilute any energy density associated with increased barotropic fluid, whereas it is conditional in the case of inflation. 

\subsection{Non-minimal frame}
Similar to previous calculations, utilizing the relations $\tilde{\gamma}$ and $\tilde{\mu}$ in terms of $\gamma$ and $\mu$ in Eqs. \eqref{eq:gammaconf} and \eqref{eq:muconf}, the Lyapunov exponents in the newly conformally modified theory can be found directly as
\begin{eqnarray}\label{eq:eigennonmin}
	\tilde{\lambda}_1 = \frac{2 (1-\alpha )+\beta  \left(1-3 w_m\right)}{\beta },\quad\tilde{\lambda}_2 = \frac{1-2 \alpha }{\beta}
\end{eqnarray}
It is clear that when we compare the above relation to the minimally coupled case (cf. Eq. 	\eqref{eq:eigenmin}), we find that
\begin{eqnarray}
	\lambda_{1} N \neq \tilde{\lambda}_1 \tilde{N},\quad \lambda_{2} N = \tilde{\lambda}_2 \tilde{N},
\end{eqnarray}
leading one to conclude that stability equivalence has been violated. Again, this is due to the fact that we left the additional matter unaffected during the conformal transformation; hence, the two theories are not conformally related. Let us now investigate the stability criteria in the new theory. Using the above relation stated in Eq. \eqref{eq:eigennonmin}, we get the criteria as
\begin{eqnarray}
	&&\text{Expansion:} \begin{cases}
		& w_m < \frac{1}{3},~ \alpha < \frac{1}{2},~ \text{and}~\beta  > \frac{2(1 - \alpha)}{(3w_m -1)}, \\
		&\text{or,}\\
		& w_m \geq \frac{1}{3}, ~\text{and}~\alpha < \frac{1}{2}.  
	\end{cases} \nonumber\\
	&& \\
	&&\text{Contraction:} \begin{cases}
		& w_m \leq \frac{1}{3},~\text{and}~\alpha < \frac{1}{2},\\
		&\text{or,}\\
		& w_m > \frac{1}{3}, ~\alpha < \frac{1}{2},~\text{and}~\beta < \frac{2(1 - \alpha)}{(3w_m -1)}.
	\end{cases}\nonumber\\
&&
\end{eqnarray}

Let's rewrite the above expression in a straightforward yet different way as

\begin{eqnarray}
	&&\alpha < \frac{1}{2}, \quad \text{and} \\
	&&\beta \begin{cases}
		>- \frac{ 2(1- \alpha)}{(1 - 3 w_m)},\quad \text{for} \quad w_m < \frac{1}{3},\\
		 < \frac{ 2(1- \alpha)}{(3 w_m - 1)},\quad ~~ \text{for} \quad w_m > \frac{1}{3}.
	\end{cases}
\end{eqnarray}
As a result, comparing this expression to the preceding one becomes easier. Again, in the minimal frame, $\alpha$ reflects the scale factor exponent, whereas in the non-minimal frame, $\beta$ does. Furthermore, as always, $\beta < 0$ denotes accelerated expansion, whereas positive $\beta$ indicates a contracting (bouncing) solution. As a result, although in the simplest situation, not all values of $\alpha$ ensure the stability requirement (cf. Eq. \eqref{eq:cond_stab_min_ex}), in this case, $\alpha$, along with $\beta$, do. Consider the state parameter equation to be in between the desired range, i.e., $-1 < w_m <1$. The stability condition in that situation is 
\begin{eqnarray}
	-\frac{(1 - \alpha)}{2} <\beta < (1 - \alpha).
\end{eqnarray}
Since $\alpha$ must meet the stringent condition of $\alpha < 1/2$ (cf. Eq. \eqref{eq:eigennonmin}), the strong condition is now 
\begin{eqnarray}
	-\frac{1}{4} < \beta < \frac{1}{2}.
\end{eqnarray} It is worth noting that this condition, coupled with $\alpha < 1/2$ and $-1 < w_m < 1$, always assures stability. In fact, the biggest achievement of non-minimal coupling is that it is always feasible to achieve stability even when $|w_m|> 1$ is chosen, i.e.,
\begin{eqnarray}
	-\frac{1}{(1 - 3w_m)}  < \beta < \frac{1}{(3w_m - 1)}.
\end{eqnarray}

As the magnitude of $w_m$ increases, the range of $\beta$ necessary for stability decreases. However, it still leaves an appropriate range of $\beta$ such that the above criterion is satisfied, and the solution is always an attractor.

We have accomplished a novel and intriguing result, thus there are few points worth mentioning. $\beta$ is determined by how we apply the conformal transformation, whereas $\alpha$ determines the model's observational signature, which is previously known \cite{Nandi:2020sif, Nandi:2020szp, Nandi:2022twa}. As a result, while there is a bound on $\alpha$ appearing immediately from the perturbation, i.e., $-1 < \alpha < 0$ is forbidden, there is no similar constraint on $\beta.$ As a result, the choice of choosing $\beta$ is absolutely random. Second, while the stability of slow-roll expansion was a condition in the minimal case, we now have tangible stability if $\beta$ is chosen correctly for the same choice of $\alpha$. That is, regardless of the choice of $w_m,$ if $\beta$ is chosen to be very close to zero (i.e., ekpyrotic expansion or contraction), the system will always find stability. For this reason, using classical perturbation analysis, we can state unequivocally that ekpyrosis (whether expansion or contraction) always dilutes perturbations, whereas inflation does not guarantee to, even in general non-minimal frames. This is the second and most significant outcome from this study.

\section{Summary and conclusions}\label{sec:6-conclu}

Let us first summarize what this article has accomplished. First, we analyzed the minimal Einstein's theory in the absence of an additional fluid and performed a dynamical analysis of the system. Only $\alpha < 1/2$ provides the stability requirement, where $\alpha$ denotes the exponent of the scale factor solution in the conformal time. Then, we conformally modified the theory so that the scale factor solution in the new theory can be written as $a(\eta) \propto (-\eta)^\beta$. Again in the non-minimal frame, we examined the stability condition and discovered that stability is equivalent, i.e., the condition is independent of $\beta$ and entirely dependent on $\alpha$ with $\alpha <  1/2.$ This is the equivalence of stability in a conformally connected frame, as previously investigated in \cite{Nandi:2019xlj}.

Next, an additional barotropic fluid was added to the minimal theory. We analyzed the stability condition once again. In addition to the one condition previously acquired, we now have a second condition that is dependent on the equation of state of the barotropic fluid, as predicted. Instantaneously, we discovered that $\alpha < 1/2$ does not always guarantee stability. Stability exists only when $\alpha$ approaches zero, i.e., ekpyrosis, whereas $\alpha < -1$ (i.e., inflation) is conditional. However, $-1 < \alpha < 0$ is ruled out because it leads to ghosts, while $\alpha > 0$ is ruled out by the observations. As a result, in minimal Einstein's theory, the inclusion of additional fluid perturbations does not ensure attractor solution.

We then moved to non-minimal theory by conformally transforming the theory, as we had done previously, with the difference that additional fluid remained unaffected by the change. This means that conformal invariance has been broken, as well as the stability equivalence. Consequently, the stability criteria have also changed, and the result is intriguing. These criteria now depend on $\alpha$, $\beta$, and $w_m$. Additionally, with the tuning of $\beta$, which is completely arbitrary and within our control to fix by conformal transformation, any value of $\alpha$ and $w_m$ can now lead to a stable solution, which is not conceivable in the minimal theory. In addition, only $\alpha$, not $\beta$, determines the observable signature of the theory, presuming that only a scalar field-dominated solution is viable for the early Universe. This is because, curvature and tensor perturbations remain invariant under conformal transformation. This invariance even holds for higher-order perturbations, as the interaction Hamiltonian at any order of perturbations (needed to evaluate for the higher-order correlation functions. See, for instance, Refs. \cite{Maldacena2003, Nandi:2015ogk, Nandi:2016pfr}) does not change under conformal transformation. As a result, even if the perturbations and thus the observables remain unchanged, the non-minimal theory becomes far more desirable than the original minimal counterpart.

In other words, using only the scalar field, one can construct a theory, such as slow-roll inflation, in Einstein's minimal frame and then transform it into a non-minimal model in which the solution is ekpyrotic. As a consequence, it is possible to assert that the new theory is more beneficial and classically stable than the original inflationary theory. These examples have already been demonstrated in Refs. \cite{Nandi:2020sif, Nandi:2020szp, Nandi:2022twa}, and the broader ramifications of such transformation are now demonstrated in this paper.

Any such result would have far-reaching effects. At first glance, such a result seems straightforward and unimportant, given that we can simply apply the conformal transformation to a single theory without influencing the barotropic fluid. However, as the early Universe era comes to an end, the impact of the barotropic fluid automatically kicks in, leaving a testable theory to be examined in the setting of the late-time Universe. Solutions other than the scalar field dominated solution, which were left out on purpose, govern these solutions and, if studied thoroughly, can have exciting and observable effects. Simple examples of these are research into the reheating era and the $H_0$ tension. In addition, as was previously noted, non-minimal stability with a de-Sitter like barotropic fluid might be seen as an answer to the naturalness of the cosmological constant problem. The reason for this is that, as the scalar field solution is very stable, the effective energy density, $\rho_m/f^2(\phi) \equiv \Lambda/f^2(\phi)$ (i.e., $\Omega_m H^2$), decreases as compared to the Hubble energy (cf. Eq. \eqref{eq:ham_constraint}). It is only after a long period of time that the effective energy density becomes identical to $\Lambda$ as a result of the decay of $f(\phi)$ to unity. One can think of this like the ``flatness problem" of the early Universe.  Third, we didn't cover many interesting but more complicated models like the non-canonical model, Galileon theory, generic Horndeski theory, etc., adhering instead to canonical scalar field theory and its conformally modified counterpart. As a fourth point, disformal transformation, like conformal transformation, renders scalar and tensor perturbations invariant, opening the door to the possibility of experimenting with disformal transformation to transform the original theory. These questions have yet to be resolved, but they hold the promise of illuminating insights and fresh takes on the universe. At the moment, we are doing a significantly more thorough investigation of these prospects.

\section*{Acknowledgements}
DN is supported by the DST, Government of India through the DST-INSPIRE Faculty fellowship (04/2020/002142). MK is supported by a DST-INSPIRE Fellowship under the reference number: IF170808, DST, Government of India. DN and MK are also very thankful to the Department of Physics and Astrophysics, University of Delhi. MK and DN also acknowledge facilities provided by the IUCAA Centre for Astronomy Research and Development (ICARD), University of Delhi.

\bibliographystyle{apsrev4-1}

\begin{thebibliography}{75}%
	\makeatletter
	\providecommand \@ifxundefined [1]{%
		\@ifx{#1\undefined}
	}%
	\providecommand \@ifnum [1]{%
		\ifnum #1\expandafter \@firstoftwo
		\else \expandafter \@secondoftwo
		\fi
	}%
	\providecommand \@ifx [1]{%
		\ifx #1\expandafter \@firstoftwo
		\else \expandafter \@secondoftwo
		\fi
	}%
	\providecommand \natexlab [1]{#1}%
	\providecommand \enquote  [1]{``#1''}%
	\providecommand \bibnamefont  [1]{#1}%
	\providecommand \bibfnamefont [1]{#1}%
	\providecommand \citenamefont [1]{#1}%
	\providecommand \href@noop [0]{\@secondoftwo}%
	\providecommand \href [0]{\begingroup \@sanitize@url \@href}%
	\providecommand \@href[1]{\@@startlink{#1}\@@href}%
	\providecommand \@@href[1]{\endgroup#1\@@endlink}%
	\providecommand \@sanitize@url [0]{\catcode `\\12\catcode `\$12\catcode
		`\&12\catcode `\#12\catcode `\^12\catcode `\_12\catcode `\%12\relax}%
	\providecommand \@@startlink[1]{}%
	\providecommand \@@endlink[0]{}%
	\providecommand \url  [0]{\begingroup\@sanitize@url \@url }%
	\providecommand \@url [1]{\endgroup\@href {#1}{\urlprefix }}%
	\providecommand \urlprefix  [0]{URL }%
	\providecommand \Eprint [0]{\href }%
	\providecommand \doibase [0]{http://dx.doi.org/}%
	\providecommand \selectlanguage [0]{\@gobble}%
	\providecommand \bibinfo  [0]{\@secondoftwo}%
	\providecommand \bibfield  [0]{\@secondoftwo}%
	\providecommand \translation [1]{[#1]}%
	\providecommand \BibitemOpen [0]{}%
	\providecommand \bibitemStop [0]{}%
	\providecommand \bibitemNoStop [0]{.\EOS\space}%
	\providecommand \EOS [0]{\spacefactor3000\relax}%
	\providecommand \BibitemShut  [1]{\csname bibitem#1\endcsname}%
	\let\auto@bib@innerbib\@empty
	\bibitem [{\citenamefont {Starobinsky}(1980)}]{STAROBINSKY198099}%
	\BibitemOpen
	\bibfield  {author} {\bibinfo {author} {\bibfnamefont {A.}~\bibnamefont
			{Starobinsky}},\ }\href {\doibase
		https://doi.org/10.1016/0370-2693(80)90670-X} {\bibfield  {journal} {\bibinfo
			{journal} {Physics Letters B}\ }\textbf {\bibinfo {volume} {91}},\ \bibinfo
		{pages} {99 } (\bibinfo {year} {1980})}\BibitemShut {NoStop}%
	\bibitem [{\citenamefont {Sato}(1981)}]{Sato:1981}%
	\BibitemOpen
	\bibfield  {author} {\bibinfo {author} {\bibfnamefont {K.}~\bibnamefont
			{Sato}},\ }\href {\doibase 10.1093/mnras/195.3.467} {\bibfield  {journal}
		{\bibinfo  {journal} {Monthly Notices of the Royal Astronomical Society}\
		}\textbf {\bibinfo {volume} {195}},\ \bibinfo {pages} {467} (\bibinfo {year}
		{1981})},\ \Eprint
	{http://arxiv.org/abs/http://oup.prod.sis.lan/mnras/article-pdf/195/3/467/4065201/mnras195-0467.pdf}
	{http://oup.prod.sis.lan/mnras/article-pdf/195/3/467/4065201/mnras195-0467.pdf}
	\BibitemShut {NoStop}%
	\bibitem [{\citenamefont {Guth}(1981)}]{Guth:1981}%
	\BibitemOpen
	\bibfield  {author} {\bibinfo {author} {\bibfnamefont {A.~H.}\ \bibnamefont
			{Guth}},\ }\href {\doibase 10.1103/PhysRevD.23.347} {\bibfield  {journal}
		{\bibinfo  {journal} {Phys. Rev. D}\ }\textbf {\bibinfo {volume} {23}},\
		\bibinfo {pages} {347} (\bibinfo {year} {1981})}\BibitemShut {NoStop}%
	\bibitem [{\citenamefont {Linde}(1982)}]{LINDE1982389}%
	\BibitemOpen
	\bibfield  {author} {\bibinfo {author} {\bibfnamefont {A.}~\bibnamefont
			{Linde}},\ }\href {\doibase https://doi.org/10.1016/0370-2693(82)91219-9}
	{\bibfield  {journal} {\bibinfo  {journal} {Physics Letters B}\ }\textbf
		{\bibinfo {volume} {108}},\ \bibinfo {pages} {389 } (\bibinfo {year}
		{1982})}\BibitemShut {NoStop}%
	\bibitem [{\citenamefont {Albrecht}\ and\ \citenamefont
		{Steinhardt}(1982)}]{Albrecht-Steinhardt:1982}%
	\BibitemOpen
	\bibfield  {author} {\bibinfo {author} {\bibfnamefont {A.}~\bibnamefont
			{Albrecht}}\ and\ \bibinfo {author} {\bibfnamefont {P.~J.}\ \bibnamefont
			{Steinhardt}},\ }\href {\doibase 10.1103/PhysRevLett.48.1220} {\bibfield
		{journal} {\bibinfo  {journal} {Phys. Rev. Lett.}\ }\textbf {\bibinfo
			{volume} {48}},\ \bibinfo {pages} {1220} (\bibinfo {year}
		{1982})}\BibitemShut {NoStop}%
	\bibitem [{\citenamefont {Linde}(1983)}]{Linde:1983gd}%
	\BibitemOpen
	\bibfield  {author} {\bibinfo {author} {\bibfnamefont {A.~D.}\ \bibnamefont
			{Linde}},\ }\href {\doibase 10.1016/0370-2693(83)90837-7} {\bibfield
		{journal} {\bibinfo  {journal} {Phys. Lett.}\ }\textbf {\bibinfo {volume}
			{129B}},\ \bibinfo {pages} {177} (\bibinfo {year} {1983})}\BibitemShut
	{NoStop}%
	\bibitem [{\citenamefont {Mukhanov}\ and\ \citenamefont
		{Chibisov}(1981)}]{Mukhanov:1981xt}%
	\BibitemOpen
	\bibfield  {author} {\bibinfo {author} {\bibfnamefont {V.~F.}\ \bibnamefont
			{Mukhanov}}\ and\ \bibinfo {author} {\bibfnamefont {G.~V.}\ \bibnamefont
			{Chibisov}},\ }\href@noop {} {\bibfield  {journal} {\bibinfo  {journal} {JETP
				Lett.}\ }\textbf {\bibinfo {volume} {33}},\ \bibinfo {pages} {532} (\bibinfo
		{year} {1981})},\ \bibinfo {note} {[Pisma Zh. Eksp. Teor.
		Fiz.33,549(1981)]}\BibitemShut {NoStop}%
	\bibitem [{\citenamefont {Hawking}(1982)}]{HAWKING1982295}%
	\BibitemOpen
	\bibfield  {author} {\bibinfo {author} {\bibfnamefont {S.}~\bibnamefont
			{Hawking}},\ }\href {\doibase https://doi.org/10.1016/0370-2693(82)90373-2}
	{\bibfield  {journal} {\bibinfo  {journal} {Physics Letters B}\ }\textbf
		{\bibinfo {volume} {115}},\ \bibinfo {pages} {295 } (\bibinfo {year}
		{1982})}\BibitemShut {NoStop}%
	\bibitem [{\citenamefont {Starobinsky}(1982)}]{STAROBINSKY1982175}%
	\BibitemOpen
	\bibfield  {author} {\bibinfo {author} {\bibfnamefont {A.}~\bibnamefont
			{Starobinsky}},\ }\href {\doibase
		http://dx.doi.org/10.1016/0370-2693(82)90541-X} {\bibfield  {journal}
		{\bibinfo  {journal} {Physics Letters B}\ }\textbf {\bibinfo {volume}
			{117}},\ \bibinfo {pages} {175 } (\bibinfo {year} {1982})}\BibitemShut
	{NoStop}%
	\bibitem [{\citenamefont {Guth}\ and\ \citenamefont {Pi}(1982)}]{Guth:1982}%
	\BibitemOpen
	\bibfield  {author} {\bibinfo {author} {\bibfnamefont {A.~H.}\ \bibnamefont
			{Guth}}\ and\ \bibinfo {author} {\bibfnamefont {S.-Y.}\ \bibnamefont {Pi}},\
	}\href {\doibase 10.1103/PhysRevLett.49.1110} {\bibfield  {journal} {\bibinfo
			{journal} {Phys. Rev. Lett.}\ }\textbf {\bibinfo {volume} {49}},\ \bibinfo
		{pages} {1110} (\bibinfo {year} {1982})}\BibitemShut {NoStop}%
	\bibitem [{\citenamefont {Vilenkin}(1983)}]{VILENKIN1983527}%
	\BibitemOpen
	\bibfield  {author} {\bibinfo {author} {\bibfnamefont {A.}~\bibnamefont
			{Vilenkin}},\ }\href {\doibase https://doi.org/10.1016/0550-3213(83)90208-0}
	{\bibfield  {journal} {\bibinfo  {journal} {Nuclear Physics B}\ }\textbf
		{\bibinfo {volume} {226}},\ \bibinfo {pages} {527 } (\bibinfo {year}
		{1983})}\BibitemShut {NoStop}%
	\bibitem [{\citenamefont {Bardeen}\ \emph {et~al.}(1983)\citenamefont
		{Bardeen}, \citenamefont {Steinhardt},\ and\ \citenamefont
		{Turner}}]{Bardeen:1983}%
	\BibitemOpen
	\bibfield  {author} {\bibinfo {author} {\bibfnamefont {J.~M.}\ \bibnamefont
			{Bardeen}}, \bibinfo {author} {\bibfnamefont {P.~J.}\ \bibnamefont
			{Steinhardt}}, \ and\ \bibinfo {author} {\bibfnamefont {M.~S.}\ \bibnamefont
			{Turner}},\ }\href {\doibase 10.1103/PhysRevD.28.679} {\bibfield  {journal}
		{\bibinfo  {journal} {Phys. Rev. D}\ }\textbf {\bibinfo {volume} {28}},\
		\bibinfo {pages} {679} (\bibinfo {year} {1983})}\BibitemShut {NoStop}%
	\bibitem [{\citenamefont {Starobinsky}(1979)}]{Starobinsky:1979ty}%
	\BibitemOpen
	\bibfield  {author} {\bibinfo {author} {\bibfnamefont {A.~A.}\ \bibnamefont
			{Starobinsky}},\ }\href@noop {} {\bibfield  {journal} {\bibinfo  {journal}
			{JETP Lett.}\ }\textbf {\bibinfo {volume} {30}},\ \bibinfo {pages} {682}
		(\bibinfo {year} {1979})}\BibitemShut {NoStop}%
	\bibitem [{\citenamefont {Akrami}\ \emph {et~al.}(2018)\citenamefont {Akrami}
		\emph {et~al.}}]{Akrami:2018odb}%
	\BibitemOpen
	\bibfield  {author} {\bibinfo {author} {\bibfnamefont {Y.}~\bibnamefont
			{Akrami}} \emph {et~al.} (\bibinfo {collaboration} {Planck}),\ }\href@noop {}
	{\  (\bibinfo {year} {2018})},\ \Eprint {http://arxiv.org/abs/1807.06211}
	{arXiv:1807.06211 [astro-ph.CO]} \BibitemShut {NoStop}%
	\bibitem [{\citenamefont {Aghanim}\ \emph {et~al.}(2018)\citenamefont {Aghanim}
		\emph {et~al.}}]{Aghanim:2018eyx}%
	\BibitemOpen
	\bibfield  {author} {\bibinfo {author} {\bibfnamefont {N.}~\bibnamefont
			{Aghanim}} \emph {et~al.} (\bibinfo {collaboration} {Planck}),\ }\href@noop
	{} {\  (\bibinfo {year} {2018})},\ \Eprint {http://arxiv.org/abs/1807.06209}
	{arXiv:1807.06209 [astro-ph.CO]} \BibitemShut {NoStop}%
	\bibitem [{\citenamefont {Copeland}\ \emph {et~al.}(1998)\citenamefont
		{Copeland}, \citenamefont {Liddle},\ and\ \citenamefont
		{Wands}}]{Copeland:1997et}%
	\BibitemOpen
	\bibfield  {author} {\bibinfo {author} {\bibfnamefont {E.~J.}\ \bibnamefont
			{Copeland}}, \bibinfo {author} {\bibfnamefont {A.~R.}\ \bibnamefont
			{Liddle}}, \ and\ \bibinfo {author} {\bibfnamefont {D.}~\bibnamefont
			{Wands}},\ }\href {\doibase 10.1103/PhysRevD.57.4686} {\bibfield  {journal}
		{\bibinfo  {journal} {Phys. Rev. D}\ }\textbf {\bibinfo {volume} {57}},\
		\bibinfo {pages} {4686} (\bibinfo {year} {1998})},\ \Eprint
	{http://arxiv.org/abs/gr-qc/9711068} {arXiv:gr-qc/9711068} \BibitemShut
	{NoStop}%
	\bibitem [{\citenamefont {Ng}\ \emph {et~al.}(2001)\citenamefont {Ng},
		\citenamefont {Nunes},\ and\ \citenamefont {Rosati}}]{Ng:2001hs}%
	\BibitemOpen
	\bibfield  {author} {\bibinfo {author} {\bibfnamefont {S.}~\bibnamefont
			{Ng}}, \bibinfo {author} {\bibfnamefont {N.}~\bibnamefont {Nunes}}, \ and\
		\bibinfo {author} {\bibfnamefont {F.}~\bibnamefont {Rosati}},\ }\href
	{\doibase 10.1103/PhysRevD.64.083510} {\bibfield  {journal} {\bibinfo
			{journal} {Phys. Rev. D}\ }\textbf {\bibinfo {volume} {64}},\ \bibinfo
		{pages} {083510} (\bibinfo {year} {2001})},\ \Eprint
	{http://arxiv.org/abs/astro-ph/0107321} {arXiv:astro-ph/0107321} \BibitemShut
	{NoStop}%
	\bibitem [{\citenamefont {Martin}\ and\ \citenamefont
		{Brandenberger}(2001)}]{Martin:2000xs}%
	\BibitemOpen
	\bibfield  {author} {\bibinfo {author} {\bibfnamefont {J.}~\bibnamefont
			{Martin}}\ and\ \bibinfo {author} {\bibfnamefont {R.~H.}\ \bibnamefont
			{Brandenberger}},\ }\href {\doibase 10.1103/PhysRevD.63.123501} {\bibfield
		{journal} {\bibinfo  {journal} {Phys. Rev. D}\ }\textbf {\bibinfo {volume}
			{63}},\ \bibinfo {pages} {123501} (\bibinfo {year} {2001})},\ \Eprint
	{http://arxiv.org/abs/hep-th/0005209} {arXiv:hep-th/0005209} \BibitemShut
	{NoStop}%
	\bibitem [{\citenamefont {Borde}\ and\ \citenamefont
		{Vilenkin}(1994)}]{Borde:1993xh}%
	\BibitemOpen
	\bibfield  {author} {\bibinfo {author} {\bibfnamefont {A.}~\bibnamefont
			{Borde}}\ and\ \bibinfo {author} {\bibfnamefont {A.}~\bibnamefont
			{Vilenkin}},\ }\href {\doibase 10.1103/PhysRevLett.72.3305} {\bibfield
		{journal} {\bibinfo  {journal} {Phys. Rev. Lett.}\ }\textbf {\bibinfo
			{volume} {72}},\ \bibinfo {pages} {3305} (\bibinfo {year} {1994})},\ \Eprint
	{http://arxiv.org/abs/gr-qc/9312022} {arXiv:gr-qc/9312022} \BibitemShut
	{NoStop}%
	\bibitem [{\citenamefont {Borde}\ \emph {et~al.}(2003)\citenamefont {Borde},
		\citenamefont {Guth},\ and\ \citenamefont {Vilenkin}}]{Borde:2001nh}%
	\BibitemOpen
	\bibfield  {author} {\bibinfo {author} {\bibfnamefont {A.}~\bibnamefont
			{Borde}}, \bibinfo {author} {\bibfnamefont {A.~H.}\ \bibnamefont {Guth}}, \
		and\ \bibinfo {author} {\bibfnamefont {A.}~\bibnamefont {Vilenkin}},\ }\href
	{\doibase 10.1103/PhysRevLett.90.151301} {\bibfield  {journal} {\bibinfo
			{journal} {Phys. Rev. Lett.}\ }\textbf {\bibinfo {volume} {90}},\ \bibinfo
		{pages} {151301} (\bibinfo {year} {2003})},\ \Eprint
	{http://arxiv.org/abs/gr-qc/0110012} {arXiv:gr-qc/0110012} \BibitemShut
	{NoStop}%
	\bibitem [{\citenamefont {Lesnefsky}\ \emph {et~al.}(2023)\citenamefont
		{Lesnefsky}, \citenamefont {Easson},\ and\ \citenamefont
		{Davies}}]{Lesnefsky:2022fen}%
	\BibitemOpen
	\bibfield  {author} {\bibinfo {author} {\bibfnamefont {J.~E.}\ \bibnamefont
			{Lesnefsky}}, \bibinfo {author} {\bibfnamefont {D.~A.}\ \bibnamefont
			{Easson}}, \ and\ \bibinfo {author} {\bibfnamefont {P.~C.~W.}\ \bibnamefont
			{Davies}},\ }\href {\doibase 10.1103/PhysRevD.107.044024} {\bibfield
		{journal} {\bibinfo  {journal} {Phys. Rev. D}\ }\textbf {\bibinfo {volume}
			{107}},\ \bibinfo {pages} {044024} (\bibinfo {year} {2023})},\ \Eprint
	{http://arxiv.org/abs/2207.00955} {arXiv:2207.00955 [gr-qc]} \BibitemShut
	{NoStop}%
	\bibitem [{\citenamefont {Martin}\ \emph {et~al.}(2011)\citenamefont {Martin},
		\citenamefont {Ringeval},\ and\ \citenamefont {Trotta}}]{Martin:2010hh}%
	\BibitemOpen
	\bibfield  {author} {\bibinfo {author} {\bibfnamefont {J.}~\bibnamefont
			{Martin}}, \bibinfo {author} {\bibfnamefont {C.}~\bibnamefont {Ringeval}}, \
		and\ \bibinfo {author} {\bibfnamefont {R.}~\bibnamefont {Trotta}},\ }\href
	{\doibase 10.1103/PhysRevD.83.063524} {\bibfield  {journal} {\bibinfo
			{journal} {Phys. Rev.}\ }\textbf {\bibinfo {volume} {D83}},\ \bibinfo {pages}
		{063524} (\bibinfo {year} {2011})},\ \Eprint {http://arxiv.org/abs/1009.4157}
	{arXiv:1009.4157 [astro-ph.CO]} \BibitemShut {NoStop}%
	\bibitem [{\citenamefont {Martin}\ \emph
		{et~al.}(2014{\natexlab{a}})\citenamefont {Martin}, \citenamefont
		{Ringeval},\ and\ \citenamefont {Vennin}}]{Martin:2013tda}%
	\BibitemOpen
	\bibfield  {author} {\bibinfo {author} {\bibfnamefont {J.}~\bibnamefont
			{Martin}}, \bibinfo {author} {\bibfnamefont {C.}~\bibnamefont {Ringeval}}, \
		and\ \bibinfo {author} {\bibfnamefont {V.}~\bibnamefont {Vennin}},\ }\href
	{\doibase 10.1016/j.dark.2014.01.003} {\bibfield  {journal} {\bibinfo
			{journal} {Phys. Dark Univ.}\ }\textbf {\bibinfo {volume} {5-6}},\ \bibinfo
		{pages} {75} (\bibinfo {year} {2014}{\natexlab{a}})},\ \Eprint
	{http://arxiv.org/abs/1303.3787} {arXiv:1303.3787 [astro-ph.CO]} \BibitemShut
	{NoStop}%
	\bibitem [{\citenamefont {Martin}\ \emph
		{et~al.}(2014{\natexlab{b}})\citenamefont {Martin}, \citenamefont {Ringeval},
		\citenamefont {Trotta},\ and\ \citenamefont {Vennin}}]{Martin:2013nzq}%
	\BibitemOpen
	\bibfield  {author} {\bibinfo {author} {\bibfnamefont {J.}~\bibnamefont
			{Martin}}, \bibinfo {author} {\bibfnamefont {C.}~\bibnamefont {Ringeval}},
		\bibinfo {author} {\bibfnamefont {R.}~\bibnamefont {Trotta}}, \ and\ \bibinfo
		{author} {\bibfnamefont {V.}~\bibnamefont {Vennin}},\ }\href {\doibase
		10.1088/1475-7516/2014/03/039} {\bibfield  {journal} {\bibinfo  {journal}
			{JCAP}\ }\textbf {\bibinfo {volume} {1403}},\ \bibinfo {pages} {039}
		(\bibinfo {year} {2014}{\natexlab{b}})},\ \Eprint
	{http://arxiv.org/abs/1312.3529} {arXiv:1312.3529 [astro-ph.CO]} \BibitemShut
	{NoStop}%
	\bibitem [{\citenamefont {Martin}\ \emph
		{et~al.}(2014{\natexlab{c}})\citenamefont {Martin}, \citenamefont
		{Ringeval},\ and\ \citenamefont {Vennin}}]{Martin:2014rqa}%
	\BibitemOpen
	\bibfield  {author} {\bibinfo {author} {\bibfnamefont {J.}~\bibnamefont
			{Martin}}, \bibinfo {author} {\bibfnamefont {C.}~\bibnamefont {Ringeval}}, \
		and\ \bibinfo {author} {\bibfnamefont {V.}~\bibnamefont {Vennin}},\ }\href
	{\doibase 10.1088/1475-7516/2014/10/038} {\bibfield  {journal} {\bibinfo
			{journal} {JCAP}\ }\textbf {\bibinfo {volume} {1410}},\ \bibinfo {pages}
		{038} (\bibinfo {year} {2014}{\natexlab{c}})},\ \Eprint
	{http://arxiv.org/abs/1407.4034} {arXiv:1407.4034 [astro-ph.CO]} \BibitemShut
	{NoStop}%
	\bibitem [{\citenamefont {Novello}\ and\ \citenamefont
		{Bergliaffa}(2008)}]{Novello:2008ra}%
	\BibitemOpen
	\bibfield  {author} {\bibinfo {author} {\bibfnamefont {M.}~\bibnamefont
			{Novello}}\ and\ \bibinfo {author} {\bibfnamefont {S.~E.~P.}\ \bibnamefont
			{Bergliaffa}},\ }\href {\doibase 10.1016/j.physrep.2008.04.006} {\bibfield
		{journal} {\bibinfo  {journal} {Phys. Rept.}\ }\textbf {\bibinfo {volume}
			{463}},\ \bibinfo {pages} {127} (\bibinfo {year} {2008})},\ \Eprint
	{http://arxiv.org/abs/0802.1634} {arXiv:0802.1634 [astro-ph]} \BibitemShut
	{NoStop}%
	\bibitem [{\citenamefont {Cai}(2014)}]{Cai:2014bea}%
	\BibitemOpen
	\bibfield  {author} {\bibinfo {author} {\bibfnamefont {Y.-F.}\ \bibnamefont
			{Cai}},\ }\href {\doibase 10.1007/s11433-014-5512-3} {\bibfield  {journal}
		{\bibinfo  {journal} {Sci. China Phys. Mech. Astron.}\ }\textbf {\bibinfo
			{volume} {57}},\ \bibinfo {pages} {1414} (\bibinfo {year} {2014})},\ \Eprint
	{http://arxiv.org/abs/1405.1369} {arXiv:1405.1369 [hep-th]} \BibitemShut
	{NoStop}%
	\bibitem [{\citenamefont {Battefeld}\ and\ \citenamefont
		{Peter}(2015)}]{Battefeld:2014uga}%
	\BibitemOpen
	\bibfield  {author} {\bibinfo {author} {\bibfnamefont {D.}~\bibnamefont
			{Battefeld}}\ and\ \bibinfo {author} {\bibfnamefont {P.}~\bibnamefont
			{Peter}},\ }\href {\doibase 10.1016/j.physrep.2014.12.004} {\bibfield
		{journal} {\bibinfo  {journal} {Phys. Rept.}\ }\textbf {\bibinfo {volume}
			{571}},\ \bibinfo {pages} {1} (\bibinfo {year} {2015})},\ \Eprint
	{http://arxiv.org/abs/1406.2790} {arXiv:1406.2790 [astro-ph.CO]} \BibitemShut
	{NoStop}%
	\bibitem [{\citenamefont {Lilley}\ and\ \citenamefont
		{Peter}(2015)}]{Lilley:2015ksa}%
	\BibitemOpen
	\bibfield  {author} {\bibinfo {author} {\bibfnamefont {M.}~\bibnamefont
			{Lilley}}\ and\ \bibinfo {author} {\bibfnamefont {P.}~\bibnamefont {Peter}},\
	}\href {\doibase 10.1016/j.crhy.2015.08.009} {\bibfield  {journal} {\bibinfo
			{journal} {Comptes Rendus Physique}\ }\textbf {\bibinfo {volume} {16}},\
		\bibinfo {pages} {1038} (\bibinfo {year} {2015})},\ \Eprint
	{http://arxiv.org/abs/1503.06578} {arXiv:1503.06578 [astro-ph.CO]}
	\BibitemShut {NoStop}%
	\bibitem [{\citenamefont {Ijjas}\ and\ \citenamefont
		{Steinhardt}(2016)}]{Ijjas:2015hcc}%
	\BibitemOpen
	\bibfield  {author} {\bibinfo {author} {\bibfnamefont {A.}~\bibnamefont
			{Ijjas}}\ and\ \bibinfo {author} {\bibfnamefont {P.~J.}\ \bibnamefont
			{Steinhardt}},\ }\href {\doibase 10.1088/0264-9381/33/4/044001} {\bibfield
		{journal} {\bibinfo  {journal} {Class. Quant. Grav.}\ }\textbf {\bibinfo
			{volume} {33}},\ \bibinfo {pages} {044001} (\bibinfo {year} {2016})},\
	\Eprint {http://arxiv.org/abs/1512.09010} {arXiv:1512.09010 [astro-ph.CO]}
	\BibitemShut {NoStop}%
	\bibitem [{\citenamefont {Brandenberger}\ and\ \citenamefont
		{Peter}(2017)}]{Brandenberger:2016vhg}%
	\BibitemOpen
	\bibfield  {author} {\bibinfo {author} {\bibfnamefont {R.}~\bibnamefont
			{Brandenberger}}\ and\ \bibinfo {author} {\bibfnamefont {P.}~\bibnamefont
			{Peter}},\ }\href {\doibase 10.1007/s10701-016-0057-0} {\bibfield  {journal}
		{\bibinfo  {journal} {Found. Phys.}\ }\textbf {\bibinfo {volume} {47}},\
		\bibinfo {pages} {797} (\bibinfo {year} {2017})},\ \Eprint
	{http://arxiv.org/abs/1603.05834} {arXiv:1603.05834 [hep-th]} \BibitemShut
	{NoStop}%
	\bibitem [{\citenamefont {Cai}\ \emph {et~al.}(2009)\citenamefont {Cai},
		\citenamefont {Xue}, \citenamefont {Brandenberger},\ and\ \citenamefont
		{Zhang}}]{Cai:2009fn}%
	\BibitemOpen
	\bibfield  {author} {\bibinfo {author} {\bibfnamefont {Y.-F.}\ \bibnamefont
			{Cai}}, \bibinfo {author} {\bibfnamefont {W.}~\bibnamefont {Xue}}, \bibinfo
		{author} {\bibfnamefont {R.}~\bibnamefont {Brandenberger}}, \ and\ \bibinfo
		{author} {\bibfnamefont {X.}~\bibnamefont {Zhang}},\ }\href {\doibase
		10.1088/1475-7516/2009/05/011} {\bibfield  {journal} {\bibinfo  {journal}
			{JCAP}\ }\textbf {\bibinfo {volume} {0905}},\ \bibinfo {pages} {011}
		(\bibinfo {year} {2009})},\ \Eprint {http://arxiv.org/abs/0903.0631}
	{arXiv:0903.0631 [astro-ph.CO]} \BibitemShut {NoStop}%
	\bibitem [{\citenamefont {Gao}\ \emph {et~al.}(2015)\citenamefont {Gao},
		\citenamefont {Lilley},\ and\ \citenamefont {Peter}}]{Gao:2014eaa}%
	\BibitemOpen
	\bibfield  {author} {\bibinfo {author} {\bibfnamefont {X.}~\bibnamefont
			{Gao}}, \bibinfo {author} {\bibfnamefont {M.}~\bibnamefont {Lilley}}, \ and\
		\bibinfo {author} {\bibfnamefont {P.}~\bibnamefont {Peter}},\ }\href
	{\doibase 10.1103/PhysRevD.91.023516} {\bibfield  {journal} {\bibinfo
			{journal} {Phys. Rev. D}\ }\textbf {\bibinfo {volume} {91}},\ \bibinfo
		{pages} {023516} (\bibinfo {year} {2015})},\ \Eprint
	{http://arxiv.org/abs/1406.4119} {arXiv:1406.4119 [gr-qc]} \BibitemShut
	{NoStop}%
	\bibitem [{\citenamefont {Gao}\ \emph {et~al.}(2014)\citenamefont {Gao},
		\citenamefont {Lilley},\ and\ \citenamefont {Peter}}]{Gao:2014hea}%
	\BibitemOpen
	\bibfield  {author} {\bibinfo {author} {\bibfnamefont {X.}~\bibnamefont
			{Gao}}, \bibinfo {author} {\bibfnamefont {M.}~\bibnamefont {Lilley}}, \ and\
		\bibinfo {author} {\bibfnamefont {P.}~\bibnamefont {Peter}},\ }\href
	{\doibase 10.1088/1475-7516/2014/07/010} {\bibfield  {journal} {\bibinfo
			{journal} {JCAP}\ }\textbf {\bibinfo {volume} {07}},\ \bibinfo {pages} {010}
		(\bibinfo {year} {2014})},\ \Eprint {http://arxiv.org/abs/1403.7958}
	{arXiv:1403.7958 [gr-qc]} \BibitemShut {NoStop}%
	\bibitem [{\citenamefont {Quintin}\ \emph {et~al.}(2015)\citenamefont
		{Quintin}, \citenamefont {Sherkatghanad}, \citenamefont {Cai},\ and\
		\citenamefont {Brandenberger}}]{Quintin:2015rta}%
	\BibitemOpen
	\bibfield  {author} {\bibinfo {author} {\bibfnamefont {J.}~\bibnamefont
			{Quintin}}, \bibinfo {author} {\bibfnamefont {Z.}~\bibnamefont
			{Sherkatghanad}}, \bibinfo {author} {\bibfnamefont {Y.-F.}\ \bibnamefont
			{Cai}}, \ and\ \bibinfo {author} {\bibfnamefont {R.~H.}\ \bibnamefont
			{Brandenberger}},\ }\href {\doibase 10.1103/PhysRevD.92.063532} {\bibfield
		{journal} {\bibinfo  {journal} {Phys. Rev.}\ }\textbf {\bibinfo {volume}
			{D92}},\ \bibinfo {pages} {063532} (\bibinfo {year} {2015})},\ \Eprint
	{http://arxiv.org/abs/1508.04141} {arXiv:1508.04141 [hep-th]} \BibitemShut
	{NoStop}%
	\bibitem [{\citenamefont {Li}\ \emph {et~al.}(2017)\citenamefont {Li},
		\citenamefont {Quintin}, \citenamefont {Wang},\ and\ \citenamefont
		{Cai}}]{Li:2016xjb}%
	\BibitemOpen
	\bibfield  {author} {\bibinfo {author} {\bibfnamefont {Y.-B.}\ \bibnamefont
			{Li}}, \bibinfo {author} {\bibfnamefont {J.}~\bibnamefont {Quintin}},
		\bibinfo {author} {\bibfnamefont {D.-G.}\ \bibnamefont {Wang}}, \ and\
		\bibinfo {author} {\bibfnamefont {Y.-F.}\ \bibnamefont {Cai}},\ }\href
	{\doibase 10.1088/1475-7516/2017/03/031} {\bibfield  {journal} {\bibinfo
			{journal} {JCAP}\ }\textbf {\bibinfo {volume} {1703}},\ \bibinfo {pages}
		{031} (\bibinfo {year} {2017})},\ \Eprint {http://arxiv.org/abs/1612.02036}
	{arXiv:1612.02036 [hep-th]} \BibitemShut {NoStop}%
	\bibitem [{\citenamefont {Akama}\ \emph {et~al.}(2020)\citenamefont {Akama},
		\citenamefont {Hirano},\ and\ \citenamefont {Kobayashi}}]{Akama:2019qeh}%
	\BibitemOpen
	\bibfield  {author} {\bibinfo {author} {\bibfnamefont {S.}~\bibnamefont
			{Akama}}, \bibinfo {author} {\bibfnamefont {S.}~\bibnamefont {Hirano}}, \
		and\ \bibinfo {author} {\bibfnamefont {T.}~\bibnamefont {Kobayashi}},\ }\href
	{\doibase 10.1103/PhysRevD.101.043529} {\bibfield  {journal} {\bibinfo
			{journal} {Phys. Rev. D}\ }\textbf {\bibinfo {volume} {101}},\ \bibinfo
		{pages} {043529} (\bibinfo {year} {2020})},\ \Eprint
	{http://arxiv.org/abs/1908.10663} {arXiv:1908.10663 [gr-qc]} \BibitemShut
	{NoStop}%
	\bibitem [{\citenamefont {Kothari}\ and\ \citenamefont
		{Nandi}(2019)}]{Kothari:2019yyw}%
	\BibitemOpen
	\bibfield  {author} {\bibinfo {author} {\bibfnamefont {R.}~\bibnamefont
			{Kothari}}\ and\ \bibinfo {author} {\bibfnamefont {D.}~\bibnamefont
			{Nandi}},\ }\href {\doibase 10.1088/1475-7516/2019/10/026} {\bibfield
		{journal} {\bibinfo  {journal} {JCAP}\ }\textbf {\bibinfo {volume} {10}},\
		\bibinfo {pages} {026} (\bibinfo {year} {2019})},\ \Eprint
	{http://arxiv.org/abs/1901.06538} {arXiv:1901.06538 [astro-ph.CO]}
	\BibitemShut {NoStop}%
	\bibitem [{\citenamefont {Kobayashi}(2016)}]{Kobayashi:2016xpl}%
	\BibitemOpen
	\bibfield  {author} {\bibinfo {author} {\bibfnamefont {T.}~\bibnamefont
			{Kobayashi}},\ }\href {\doibase 10.1103/PhysRevD.94.043511} {\bibfield
		{journal} {\bibinfo  {journal} {Phys. Rev.}\ }\textbf {\bibinfo {volume}
			{D94}},\ \bibinfo {pages} {043511} (\bibinfo {year} {2016})},\ \Eprint
	{http://arxiv.org/abs/1606.05831} {arXiv:1606.05831 [hep-th]} \BibitemShut
	{NoStop}%
	\bibitem [{\citenamefont {Libanov}\ \emph {et~al.}(2016)\citenamefont
		{Libanov}, \citenamefont {Mironov},\ and\ \citenamefont
		{Rubakov}}]{Libanov:2016kfc}%
	\BibitemOpen
	\bibfield  {author} {\bibinfo {author} {\bibfnamefont {M.}~\bibnamefont
			{Libanov}}, \bibinfo {author} {\bibfnamefont {S.}~\bibnamefont {Mironov}}, \
		and\ \bibinfo {author} {\bibfnamefont {V.}~\bibnamefont {Rubakov}},\ }\href
	{\doibase 10.1088/1475-7516/2016/08/037} {\bibfield  {journal} {\bibinfo
			{journal} {JCAP}\ }\textbf {\bibinfo {volume} {1608}},\ \bibinfo {pages}
		{037} (\bibinfo {year} {2016})},\ \Eprint {http://arxiv.org/abs/1605.05992}
	{arXiv:1605.05992 [hep-th]} \BibitemShut {NoStop}%
	\bibitem [{\citenamefont {Ijjas}\ and\ \citenamefont
		{Steinhardt}(2017)}]{Ijjas:2016vtq}%
	\BibitemOpen
	\bibfield  {author} {\bibinfo {author} {\bibfnamefont {A.}~\bibnamefont
			{Ijjas}}\ and\ \bibinfo {author} {\bibfnamefont {P.~J.}\ \bibnamefont
			{Steinhardt}},\ }\href {\doibase 10.1016/j.physletb.2016.11.047} {\bibfield
		{journal} {\bibinfo  {journal} {Phys. Lett.}\ }\textbf {\bibinfo {volume}
			{B764}},\ \bibinfo {pages} {289} (\bibinfo {year} {2017})},\ \Eprint
	{http://arxiv.org/abs/1609.01253} {arXiv:1609.01253 [gr-qc]} \BibitemShut
	{NoStop}%
	\bibitem [{\citenamefont {Banerjee}\ \emph {et~al.}(2019)\citenamefont
		{Banerjee}, \citenamefont {Cai},\ and\ \citenamefont
		{Saridakis}}]{Banerjee:2018svi}%
	\BibitemOpen
	\bibfield  {author} {\bibinfo {author} {\bibfnamefont {S.}~\bibnamefont
			{Banerjee}}, \bibinfo {author} {\bibfnamefont {Y.-F.}\ \bibnamefont {Cai}}, \
		and\ \bibinfo {author} {\bibfnamefont {E.~N.}\ \bibnamefont {Saridakis}},\
	}\href {\doibase 10.1088/1361-6382/ab256a} {\bibfield  {journal} {\bibinfo
			{journal} {Class. Quant. Grav.}\ }\textbf {\bibinfo {volume} {36}},\ \bibinfo
		{pages} {135009} (\bibinfo {year} {2019})},\ \Eprint
	{http://arxiv.org/abs/1808.01170} {arXiv:1808.01170 [gr-qc]} \BibitemShut
	{NoStop}%
	\bibitem [{\citenamefont {Cai}\ \emph {et~al.}(2017)\citenamefont {Cai},
		\citenamefont {Wan}, \citenamefont {Li}, \citenamefont {Qiu},\ and\
		\citenamefont {Piao}}]{Cai:2016thi}%
	\BibitemOpen
	\bibfield  {author} {\bibinfo {author} {\bibfnamefont {Y.}~\bibnamefont
			{Cai}}, \bibinfo {author} {\bibfnamefont {Y.}~\bibnamefont {Wan}}, \bibinfo
		{author} {\bibfnamefont {H.-G.}\ \bibnamefont {Li}}, \bibinfo {author}
		{\bibfnamefont {T.}~\bibnamefont {Qiu}}, \ and\ \bibinfo {author}
		{\bibfnamefont {Y.-S.}\ \bibnamefont {Piao}},\ }\href {\doibase
		10.1007/JHEP01(2017)090} {\bibfield  {journal} {\bibinfo  {journal} {JHEP}\
		}\textbf {\bibinfo {volume} {01}},\ \bibinfo {pages} {090} (\bibinfo {year}
		{2017})},\ \Eprint {http://arxiv.org/abs/1610.03400} {arXiv:1610.03400
		[gr-qc]} \BibitemShut {NoStop}%
	\bibitem [{\citenamefont {Cai}\ and\ \citenamefont {Piao}(2017)}]{Cai:2017dyi}%
	\BibitemOpen
	\bibfield  {author} {\bibinfo {author} {\bibfnamefont {Y.}~\bibnamefont
			{Cai}}\ and\ \bibinfo {author} {\bibfnamefont {Y.-S.}\ \bibnamefont {Piao}},\
	}\href {\doibase 10.1007/JHEP09(2017)027} {\bibfield  {journal} {\bibinfo
			{journal} {JHEP}\ }\textbf {\bibinfo {volume} {09}},\ \bibinfo {pages} {027}
		(\bibinfo {year} {2017})},\ \Eprint {http://arxiv.org/abs/1705.03401}
	{arXiv:1705.03401 [gr-qc]} \BibitemShut {NoStop}%
	\bibitem [{\citenamefont {Kolevatov}\ \emph {et~al.}(2017)\citenamefont
		{Kolevatov}, \citenamefont {Mironov}, \citenamefont {Sukhov},\ and\
		\citenamefont {Volkova}}]{Kolevatov:2017voe}%
	\BibitemOpen
	\bibfield  {author} {\bibinfo {author} {\bibfnamefont {R.}~\bibnamefont
			{Kolevatov}}, \bibinfo {author} {\bibfnamefont {S.}~\bibnamefont {Mironov}},
		\bibinfo {author} {\bibfnamefont {N.}~\bibnamefont {Sukhov}}, \ and\ \bibinfo
		{author} {\bibfnamefont {V.}~\bibnamefont {Volkova}},\ }\href {\doibase
		10.1088/1475-7516/2017/08/038} {\bibfield  {journal} {\bibinfo  {journal}
			{JCAP}\ }\textbf {\bibinfo {volume} {1708}},\ \bibinfo {pages} {038}
		(\bibinfo {year} {2017})},\ \Eprint {http://arxiv.org/abs/1705.06626}
	{arXiv:1705.06626 [hep-th]} \BibitemShut {NoStop}%
	\bibitem [{\citenamefont {Mironov}\ \emph {et~al.}(2018)\citenamefont
		{Mironov}, \citenamefont {Rubakov},\ and\ \citenamefont
		{Volkova}}]{Mironov:2018oec}%
	\BibitemOpen
	\bibfield  {author} {\bibinfo {author} {\bibfnamefont {S.}~\bibnamefont
			{Mironov}}, \bibinfo {author} {\bibfnamefont {V.}~\bibnamefont {Rubakov}}, \
		and\ \bibinfo {author} {\bibfnamefont {V.}~\bibnamefont {Volkova}},\ }\href
	{\doibase 10.1088/1475-7516/2018/10/050} {\bibfield  {journal} {\bibinfo
			{journal} {JCAP}\ }\textbf {\bibinfo {volume} {1810}},\ \bibinfo {pages}
		{050} (\bibinfo {year} {2018})},\ \Eprint {http://arxiv.org/abs/1807.08361}
	{arXiv:1807.08361 [hep-th]} \BibitemShut {NoStop}%
	\bibitem [{\citenamefont {Easson}\ \emph {et~al.}(2011)\citenamefont {Easson},
		\citenamefont {Sawicki},\ and\ \citenamefont {Vikman}}]{Easson:2011zy}%
	\BibitemOpen
	\bibfield  {author} {\bibinfo {author} {\bibfnamefont {D.~A.}\ \bibnamefont
			{Easson}}, \bibinfo {author} {\bibfnamefont {I.}~\bibnamefont {Sawicki}}, \
		and\ \bibinfo {author} {\bibfnamefont {A.}~\bibnamefont {Vikman}},\ }\href
	{\doibase 10.1088/1475-7516/2011/11/021} {\bibfield  {journal} {\bibinfo
			{journal} {JCAP}\ }\textbf {\bibinfo {volume} {11}},\ \bibinfo {pages} {021}
		(\bibinfo {year} {2011})},\ \Eprint {http://arxiv.org/abs/1109.1047}
	{arXiv:1109.1047 [hep-th]} \BibitemShut {NoStop}%
	\bibitem [{\citenamefont {Sawicki}\ and\ \citenamefont
		{Vikman}(2013)}]{Sawicki:2012pz}%
	\BibitemOpen
	\bibfield  {author} {\bibinfo {author} {\bibfnamefont {I.}~\bibnamefont
			{Sawicki}}\ and\ \bibinfo {author} {\bibfnamefont {A.}~\bibnamefont
			{Vikman}},\ }\href {\doibase 10.1103/PhysRevD.87.067301} {\bibfield
		{journal} {\bibinfo  {journal} {Phys. Rev. D}\ }\textbf {\bibinfo {volume}
			{87}},\ \bibinfo {pages} {067301} (\bibinfo {year} {2013})},\ \Eprint
	{http://arxiv.org/abs/1209.2961} {arXiv:1209.2961 [astro-ph.CO]} \BibitemShut
	{NoStop}%
	\bibitem [{\citenamefont {Belinskii}\ \emph {et~al.}(1970)\citenamefont
		{Belinskii}, \citenamefont {Khalatnikov},\ and\ \citenamefont
		{Lifshitz}}]{doi:10.1080/00018737000101171}%
	\BibitemOpen
	\bibfield  {author} {\bibinfo {author} {\bibfnamefont {V.}~\bibnamefont
			{Belinskii}}, \bibinfo {author} {\bibfnamefont {I.}~\bibnamefont
			{Khalatnikov}}, \ and\ \bibinfo {author} {\bibfnamefont {E.}~\bibnamefont
			{Lifshitz}},\ }\href {\doibase 10.1080/00018737000101171} {\bibfield
		{journal} {\bibinfo  {journal} {Advances in Physics}\ }\textbf {\bibinfo
			{volume} {19}},\ \bibinfo {pages} {525} (\bibinfo {year} {1970})}\BibitemShut
	{NoStop}%
	\bibitem [{\citenamefont {Karouby}\ and\ \citenamefont
		{Brandenberger}(2010)}]{Karouby:2010wt}%
	\BibitemOpen
	\bibfield  {author} {\bibinfo {author} {\bibfnamefont {J.}~\bibnamefont
			{Karouby}}\ and\ \bibinfo {author} {\bibfnamefont {R.}~\bibnamefont
			{Brandenberger}},\ }\href {\doibase 10.1103/PhysRevD.82.063532} {\bibfield
		{journal} {\bibinfo  {journal} {Phys. Rev. D}\ }\textbf {\bibinfo {volume}
			{82}},\ \bibinfo {pages} {063532} (\bibinfo {year} {2010})},\ \Eprint
	{http://arxiv.org/abs/1004.4947} {arXiv:1004.4947 [hep-th]} \BibitemShut
	{NoStop}%
	\bibitem [{\citenamefont {Karouby}\ \emph {et~al.}(2011)\citenamefont
		{Karouby}, \citenamefont {Qiu},\ and\ \citenamefont
		{Brandenberger}}]{Karouby:2011wj}%
	\BibitemOpen
	\bibfield  {author} {\bibinfo {author} {\bibfnamefont {J.}~\bibnamefont
			{Karouby}}, \bibinfo {author} {\bibfnamefont {T.}~\bibnamefont {Qiu}}, \ and\
		\bibinfo {author} {\bibfnamefont {R.}~\bibnamefont {Brandenberger}},\ }\href
	{\doibase 10.1103/PhysRevD.84.043505} {\bibfield  {journal} {\bibinfo
			{journal} {Phys. Rev. D}\ }\textbf {\bibinfo {volume} {84}},\ \bibinfo
		{pages} {043505} (\bibinfo {year} {2011})},\ \Eprint
	{http://arxiv.org/abs/1104.3193} {arXiv:1104.3193 [hep-th]} \BibitemShut
	{NoStop}%
	\bibitem [{\citenamefont {Bhattacharya}\ \emph {et~al.}(2013)\citenamefont
		{Bhattacharya}, \citenamefont {Cai},\ and\ \citenamefont
		{Das}}]{Bhattacharya:2013ut}%
	\BibitemOpen
	\bibfield  {author} {\bibinfo {author} {\bibfnamefont {K.}~\bibnamefont
			{Bhattacharya}}, \bibinfo {author} {\bibfnamefont {Y.-F.}\ \bibnamefont
			{Cai}}, \ and\ \bibinfo {author} {\bibfnamefont {S.}~\bibnamefont {Das}},\
	}\href {\doibase 10.1103/PhysRevD.87.083511} {\bibfield  {journal} {\bibinfo
			{journal} {Phys. Rev. D}\ }\textbf {\bibinfo {volume} {87}},\ \bibinfo
		{pages} {083511} (\bibinfo {year} {2013})},\ \Eprint
	{http://arxiv.org/abs/1301.0661} {arXiv:1301.0661 [hep-th]} \BibitemShut
	{NoStop}%
	\bibitem [{\citenamefont {Cai}\ \emph {et~al.}(2013)\citenamefont {Cai},
		\citenamefont {Brandenberger},\ and\ \citenamefont {Peter}}]{Cai:2013vm}%
	\BibitemOpen
	\bibfield  {author} {\bibinfo {author} {\bibfnamefont {Y.-F.}\ \bibnamefont
			{Cai}}, \bibinfo {author} {\bibfnamefont {R.}~\bibnamefont {Brandenberger}},
		\ and\ \bibinfo {author} {\bibfnamefont {P.}~\bibnamefont {Peter}},\ }\href
	{\doibase 10.1088/0264-9381/30/7/075019} {\bibfield  {journal} {\bibinfo
			{journal} {Class. Quant. Grav.}\ }\textbf {\bibinfo {volume} {30}},\ \bibinfo
		{pages} {075019} (\bibinfo {year} {2013})},\ \Eprint
	{http://arxiv.org/abs/1301.4703} {arXiv:1301.4703 [gr-qc]} \BibitemShut
	{NoStop}%
	\bibitem [{\citenamefont {Ganguly}\ and\ \citenamefont
		{Quintin}(2022)}]{Ganguly:2021pke}%
	\BibitemOpen
	\bibfield  {author} {\bibinfo {author} {\bibfnamefont {C.}~\bibnamefont
			{Ganguly}}\ and\ \bibinfo {author} {\bibfnamefont {J.}~\bibnamefont
			{Quintin}},\ }\href {\doibase 10.1103/PhysRevD.105.023532} {\bibfield
		{journal} {\bibinfo  {journal} {Phys. Rev. D}\ }\textbf {\bibinfo {volume}
			{105}},\ \bibinfo {pages} {023532} (\bibinfo {year} {2022})},\ \Eprint
	{http://arxiv.org/abs/2109.11701} {arXiv:2109.11701 [gr-qc]} \BibitemShut
	{NoStop}%
	\bibitem [{\citenamefont {Horndeski}(1974)}]{Horndeski:1974wa}%
	\BibitemOpen
	\bibfield  {author} {\bibinfo {author} {\bibfnamefont {G.~W.}\ \bibnamefont
			{Horndeski}},\ }\href {\doibase 10.1007/BF01807638} {\bibfield  {journal}
		{\bibinfo  {journal} {Int. J. Theor. Phys.}\ }\textbf {\bibinfo {volume}
			{10}},\ \bibinfo {pages} {363} (\bibinfo {year} {1974})}\BibitemShut
	{NoStop}%
	\bibitem [{\citenamefont {Gleyzes}\ \emph {et~al.}(2015)\citenamefont
		{Gleyzes}, \citenamefont {Langlois}, \citenamefont {Piazza},\ and\
		\citenamefont {Vernizzi}}]{Gleyzes:2014dya}%
	\BibitemOpen
	\bibfield  {author} {\bibinfo {author} {\bibfnamefont {J.}~\bibnamefont
			{Gleyzes}}, \bibinfo {author} {\bibfnamefont {D.}~\bibnamefont {Langlois}},
		\bibinfo {author} {\bibfnamefont {F.}~\bibnamefont {Piazza}}, \ and\ \bibinfo
		{author} {\bibfnamefont {F.}~\bibnamefont {Vernizzi}},\ }\href {\doibase
		10.1103/PhysRevLett.114.211101} {\bibfield  {journal} {\bibinfo  {journal}
			{Phys. Rev. Lett.}\ }\textbf {\bibinfo {volume} {114}},\ \bibinfo {pages}
		{211101} (\bibinfo {year} {2015})},\ \Eprint {http://arxiv.org/abs/1404.6495}
	{arXiv:1404.6495 [hep-th]} \BibitemShut {NoStop}%
	\bibitem [{\citenamefont {Kobayashi}(2019)}]{Kobayashi:2019hrl}%
	\BibitemOpen
	\bibfield  {author} {\bibinfo {author} {\bibfnamefont {T.}~\bibnamefont
			{Kobayashi}},\ }\href@noop {} {\  (\bibinfo {year} {2019})},\ \Eprint
	{http://arxiv.org/abs/1901.07183} {arXiv:1901.07183 [gr-qc]} \BibitemShut
	{NoStop}%
	\bibitem [{\citenamefont {Cai}\ \emph {et~al.}(2012)\citenamefont {Cai},
		\citenamefont {Easson},\ and\ \citenamefont {Brandenberger}}]{Cai:2012va}%
	\BibitemOpen
	\bibfield  {author} {\bibinfo {author} {\bibfnamefont {Y.-F.}\ \bibnamefont
			{Cai}}, \bibinfo {author} {\bibfnamefont {D.~A.}\ \bibnamefont {Easson}}, \
		and\ \bibinfo {author} {\bibfnamefont {R.}~\bibnamefont {Brandenberger}},\
	}\href {\doibase 10.1088/1475-7516/2012/08/020} {\bibfield  {journal}
		{\bibinfo  {journal} {JCAP}\ }\textbf {\bibinfo {volume} {08}},\ \bibinfo
		{pages} {020} (\bibinfo {year} {2012})},\ \Eprint
	{http://arxiv.org/abs/1206.2382} {arXiv:1206.2382 [hep-th]} \BibitemShut
	{NoStop}%
	\bibitem [{\citenamefont {Ilyas}\ \emph {et~al.}(2020)\citenamefont {Ilyas},
		\citenamefont {Zhu}, \citenamefont {Zheng}, \citenamefont {Cai},\ and\
		\citenamefont {Saridakis}}]{Ilyas:2020qja}%
	\BibitemOpen
	\bibfield  {author} {\bibinfo {author} {\bibfnamefont {A.}~\bibnamefont
			{Ilyas}}, \bibinfo {author} {\bibfnamefont {M.}~\bibnamefont {Zhu}}, \bibinfo
		{author} {\bibfnamefont {Y.}~\bibnamefont {Zheng}}, \bibinfo {author}
		{\bibfnamefont {Y.-F.}\ \bibnamefont {Cai}}, \ and\ \bibinfo {author}
		{\bibfnamefont {E.~N.}\ \bibnamefont {Saridakis}},\ }\href@noop {} {\
		(\bibinfo {year} {2020})},\ \Eprint {http://arxiv.org/abs/2002.08269}
	{arXiv:2002.08269 [gr-qc]} \BibitemShut {NoStop}%
	\bibitem [{\citenamefont {Dobre}\ \emph {et~al.}(2018)\citenamefont {Dobre},
		\citenamefont {Frolov}, \citenamefont {Ghersi}, \citenamefont {Ramazanov},\
		and\ \citenamefont {Vikman}}]{Dobre:2017pnt}%
	\BibitemOpen
	\bibfield  {author} {\bibinfo {author} {\bibfnamefont {D.~A.}\ \bibnamefont
			{Dobre}}, \bibinfo {author} {\bibfnamefont {A.~V.}\ \bibnamefont {Frolov}},
		\bibinfo {author} {\bibfnamefont {J.~T.~G.}\ \bibnamefont {Ghersi}}, \bibinfo
		{author} {\bibfnamefont {S.}~\bibnamefont {Ramazanov}}, \ and\ \bibinfo
		{author} {\bibfnamefont {A.}~\bibnamefont {Vikman}},\ }\href {\doibase
		10.1088/1475-7516/2018/03/020} {\bibfield  {journal} {\bibinfo  {journal}
			{JCAP}\ }\textbf {\bibinfo {volume} {03}},\ \bibinfo {pages} {020} (\bibinfo
		{year} {2018})},\ \Eprint {http://arxiv.org/abs/1712.10272} {arXiv:1712.10272
		[gr-qc]} \BibitemShut {NoStop}%
	\bibitem [{\citenamefont {Zhu}\ \emph {et~al.}(2021)\citenamefont {Zhu},
		\citenamefont {Ilyas}, \citenamefont {Zheng}, \citenamefont {Cai},\ and\
		\citenamefont {Saridakis}}]{Zhu:2021whu}%
	\BibitemOpen
	\bibfield  {author} {\bibinfo {author} {\bibfnamefont {M.}~\bibnamefont
			{Zhu}}, \bibinfo {author} {\bibfnamefont {A.}~\bibnamefont {Ilyas}}, \bibinfo
		{author} {\bibfnamefont {Y.}~\bibnamefont {Zheng}}, \bibinfo {author}
		{\bibfnamefont {Y.-F.}\ \bibnamefont {Cai}}, \ and\ \bibinfo {author}
		{\bibfnamefont {E.~N.}\ \bibnamefont {Saridakis}},\ }\href {\doibase
		10.1088/1475-7516/2021/11/045} {\bibfield  {journal} {\bibinfo  {journal}
			{JCAP}\ }\textbf {\bibinfo {volume} {11}},\ \bibinfo {pages} {045} (\bibinfo
		{year} {2021})},\ \Eprint {http://arxiv.org/abs/2108.01339} {arXiv:2108.01339
		[gr-qc]} \BibitemShut {NoStop}%
	\bibitem [{\citenamefont {Levy}\ \emph {et~al.}(2015)\citenamefont {Levy},
		\citenamefont {Ijjas},\ and\ \citenamefont {Steinhardt}}]{Levy:2015awa}%
	\BibitemOpen
	\bibfield  {author} {\bibinfo {author} {\bibfnamefont {A.~M.}\ \bibnamefont
			{Levy}}, \bibinfo {author} {\bibfnamefont {A.}~\bibnamefont {Ijjas}}, \ and\
		\bibinfo {author} {\bibfnamefont {P.~J.}\ \bibnamefont {Steinhardt}},\ }\href
	{\doibase 10.1103/PhysRevD.92.063524} {\bibfield  {journal} {\bibinfo
			{journal} {Phys. Rev.}\ }\textbf {\bibinfo {volume} {D92}},\ \bibinfo {pages}
		{063524} (\bibinfo {year} {2015})},\ \Eprint
	{http://arxiv.org/abs/1506.01011} {arXiv:1506.01011 [astro-ph.CO]}
	\BibitemShut {NoStop}%
	\bibitem [{\citenamefont {East}\ \emph {et~al.}(2016)\citenamefont {East},
		\citenamefont {Kleban}, \citenamefont {Linde},\ and\ \citenamefont
		{Senatore}}]{East:2015ggf}%
	\BibitemOpen
	\bibfield  {author} {\bibinfo {author} {\bibfnamefont {W.~E.}\ \bibnamefont
			{East}}, \bibinfo {author} {\bibfnamefont {M.}~\bibnamefont {Kleban}},
		\bibinfo {author} {\bibfnamefont {A.}~\bibnamefont {Linde}}, \ and\ \bibinfo
		{author} {\bibfnamefont {L.}~\bibnamefont {Senatore}},\ }\href {\doibase
		10.1088/1475-7516/2016/09/010} {\bibfield  {journal} {\bibinfo  {journal}
			{JCAP}\ }\textbf {\bibinfo {volume} {1609}},\ \bibinfo {pages} {010}
		(\bibinfo {year} {2016})},\ \Eprint {http://arxiv.org/abs/1511.05143}
	{arXiv:1511.05143 [hep-th]} \BibitemShut {NoStop}%
	\bibitem [{\citenamefont {Clough}\ \emph {et~al.}(2017)\citenamefont {Clough},
		\citenamefont {Lim}, \citenamefont {DiNunno}, \citenamefont {Fischler},
		\citenamefont {Flauger},\ and\ \citenamefont {Paban}}]{Clough:2016ymm}%
	\BibitemOpen
	\bibfield  {author} {\bibinfo {author} {\bibfnamefont {K.}~\bibnamefont
			{Clough}}, \bibinfo {author} {\bibfnamefont {E.~A.}\ \bibnamefont {Lim}},
		\bibinfo {author} {\bibfnamefont {B.~S.}\ \bibnamefont {DiNunno}}, \bibinfo
		{author} {\bibfnamefont {W.}~\bibnamefont {Fischler}}, \bibinfo {author}
		{\bibfnamefont {R.}~\bibnamefont {Flauger}}, \ and\ \bibinfo {author}
		{\bibfnamefont {S.}~\bibnamefont {Paban}},\ }\href {\doibase
		10.1088/1475-7516/2017/09/025} {\bibfield  {journal} {\bibinfo  {journal}
			{JCAP}\ }\textbf {\bibinfo {volume} {1709}},\ \bibinfo {pages} {025}
		(\bibinfo {year} {2017})},\ \Eprint {http://arxiv.org/abs/1608.04408}
	{arXiv:1608.04408 [hep-th]} \BibitemShut {NoStop}%
	\bibitem [{\citenamefont {Garfinkle}\ \emph {et~al.}(2008)\citenamefont
		{Garfinkle}, \citenamefont {Lim}, \citenamefont {Pretorius},\ and\
		\citenamefont {Steinhardt}}]{Garfinkle:2008ei}%
	\BibitemOpen
	\bibfield  {author} {\bibinfo {author} {\bibfnamefont {D.}~\bibnamefont
			{Garfinkle}}, \bibinfo {author} {\bibfnamefont {W.~C.}\ \bibnamefont {Lim}},
		\bibinfo {author} {\bibfnamefont {F.}~\bibnamefont {Pretorius}}, \ and\
		\bibinfo {author} {\bibfnamefont {P.~J.}\ \bibnamefont {Steinhardt}},\ }\href
	{\doibase 10.1103/PhysRevD.78.083537} {\bibfield  {journal} {\bibinfo
			{journal} {Phys. Rev.}\ }\textbf {\bibinfo {volume} {D78}},\ \bibinfo {pages}
		{083537} (\bibinfo {year} {2008})},\ \Eprint {http://arxiv.org/abs/0808.0542}
	{arXiv:0808.0542 [hep-th]} \BibitemShut {NoStop}%
	\bibitem [{\citenamefont {Nandi}(2018)}]{Nandi:2018ooh}%
	\BibitemOpen
	\bibfield  {author} {\bibinfo {author} {\bibfnamefont {D.}~\bibnamefont
			{Nandi}},\ }\href {\doibase 10.1088/1475-7516/2019/05/040} {\bibfield
		{journal} {\bibinfo  {journal} {JCAP}\ }\textbf {\bibinfo {volume} {1905}},\
		\bibinfo {pages} {040} (\bibinfo {year} {2018})},\ \Eprint
	{http://arxiv.org/abs/1811.09625} {arXiv:1811.09625 [gr-qc]} \BibitemShut
	{NoStop}%
	\bibitem [{\citenamefont {Nandi}(2019)}]{Nandi:2019xlj}%
	\BibitemOpen
	\bibfield  {author} {\bibinfo {author} {\bibfnamefont {D.}~\bibnamefont
			{Nandi}},\ }\href {\doibase 10.1103/PhysRevD.99.103532} {\bibfield  {journal}
		{\bibinfo  {journal} {Phys. Rev.}\ }\textbf {\bibinfo {volume} {D99}},\
		\bibinfo {pages} {103532} (\bibinfo {year} {2019})},\ \Eprint
	{http://arxiv.org/abs/1904.00153} {arXiv:1904.00153 [gr-qc]} \BibitemShut
	{NoStop}%
	\bibitem [{\citenamefont {Nandi}\ and\ \citenamefont
		{Sriramkumar}(2020)}]{Nandi:2019xag}%
	\BibitemOpen
	\bibfield  {author} {\bibinfo {author} {\bibfnamefont {D.}~\bibnamefont
			{Nandi}}\ and\ \bibinfo {author} {\bibfnamefont {L.}~\bibnamefont
			{Sriramkumar}},\ }\href {\doibase 10.1103/PhysRevD.101.043506} {\bibfield
		{journal} {\bibinfo  {journal} {Phys. Rev.}\ }\textbf {\bibinfo {volume}
			{D101}},\ \bibinfo {pages} {043506} (\bibinfo {year} {2020})},\ \Eprint
	{http://arxiv.org/abs/1904.13254} {arXiv:1904.13254 [gr-qc]} \BibitemShut
	{NoStop}%
	\bibitem [{\citenamefont {Nandi}(2020)}]{Nandi:2020sif}%
	\BibitemOpen
	\bibfield  {author} {\bibinfo {author} {\bibfnamefont {D.}~\bibnamefont
			{Nandi}},\ }\href {\doibase 10.1016/j.physletb.2020.135695} {\bibfield
		{journal} {\bibinfo  {journal} {Phys. Lett. B}\ }\textbf {\bibinfo {volume}
			{809}},\ \bibinfo {pages} {135695} (\bibinfo {year} {2020})},\ \Eprint
	{http://arxiv.org/abs/2003.02066} {arXiv:2003.02066 [astro-ph.CO]}
	\BibitemShut {NoStop}%
	\bibitem [{\citenamefont {Nandi}(2021)}]{Nandi:2020szp}%
	\BibitemOpen
	\bibfield  {author} {\bibinfo {author} {\bibfnamefont {D.}~\bibnamefont
			{Nandi}},\ }\href {\doibase 10.3390/universe7030062} {\bibfield  {journal}
		{\bibinfo  {journal} {Universe}\ }\textbf {\bibinfo {volume} {7}},\ \bibinfo
		{pages} {62} (\bibinfo {year} {2021})},\ \Eprint
	{http://arxiv.org/abs/2009.03134} {arXiv:2009.03134 [gr-qc]} \BibitemShut
	{NoStop}%
	\bibitem [{\citenamefont {Nandi}\ and\ \citenamefont
		{Kaur}(2022)}]{Nandi:2022twa}%
	\BibitemOpen
	\bibfield  {author} {\bibinfo {author} {\bibfnamefont {D.}~\bibnamefont
			{Nandi}}\ and\ \bibinfo {author} {\bibfnamefont {M.}~\bibnamefont {Kaur}},\
	}\href@noop {} {\  (\bibinfo {year} {2022})},\ \Eprint
	{http://arxiv.org/abs/2206.08335} {arXiv:2206.08335 [astro-ph.CO]}
	\BibitemShut {NoStop}%
	\bibitem [{\citenamefont {Copeland}\ \emph {et~al.}(2006)\citenamefont
		{Copeland}, \citenamefont {Sami},\ and\ \citenamefont
		{Tsujikawa}}]{Copeland:2006wr}%
	\BibitemOpen
	\bibfield  {author} {\bibinfo {author} {\bibfnamefont {E.~J.}\ \bibnamefont
			{Copeland}}, \bibinfo {author} {\bibfnamefont {M.}~\bibnamefont {Sami}}, \
		and\ \bibinfo {author} {\bibfnamefont {S.}~\bibnamefont {Tsujikawa}},\ }\href
	{\doibase 10.1142/S021827180600942X} {\bibfield  {journal} {\bibinfo
			{journal} {Int. J. Mod. Phys. D}\ }\textbf {\bibinfo {volume} {15}},\
		\bibinfo {pages} {1753} (\bibinfo {year} {2006})},\ \Eprint
	{http://arxiv.org/abs/hep-th/0603057} {arXiv:hep-th/0603057} \BibitemShut
	{NoStop}%
	\bibitem [{\citenamefont {Maldacena}(2003)}]{Maldacena2003}%
	\BibitemOpen
	\bibfield  {author} {\bibinfo {author} {\bibfnamefont {J.~M.}\ \bibnamefont
			{Maldacena}},\ }\href {\doibase 10.1088/1126-6708/2003/05/013} {\bibfield
		{journal} {\bibinfo  {journal} {JHEP}\ }\textbf {\bibinfo {volume} {0305}},\
		\bibinfo {pages} {013} (\bibinfo {year} {2003})},\ \Eprint
	{http://arxiv.org/abs/astro-ph/0210603} {arXiv:astro-ph/0210603 [astro-ph]}
	\BibitemShut {NoStop}%
	\bibitem [{\citenamefont {Nandi}\ and\ \citenamefont
		{Shankaranarayanan}(2016{\natexlab{a}})}]{Nandi:2015ogk}%
	\BibitemOpen
	\bibfield  {author} {\bibinfo {author} {\bibfnamefont {D.}~\bibnamefont
			{Nandi}}\ and\ \bibinfo {author} {\bibfnamefont {S.}~\bibnamefont
			{Shankaranarayanan}},\ }\href {\doibase 10.1088/1475-7516/2016/06/038}
	{\bibfield  {journal} {\bibinfo  {journal} {JCAP}\ }\textbf {\bibinfo
			{volume} {1606}},\ \bibinfo {pages} {038} (\bibinfo {year}
		{2016}{\natexlab{a}})},\ \Eprint {http://arxiv.org/abs/1512.02539}
	{arXiv:1512.02539 [gr-qc]} \BibitemShut {NoStop}%
	\bibitem [{\citenamefont {Nandi}\ and\ \citenamefont
		{Shankaranarayanan}(2016{\natexlab{b}})}]{Nandi:2016pfr}%
	\BibitemOpen
	\bibfield  {author} {\bibinfo {author} {\bibfnamefont {D.}~\bibnamefont
			{Nandi}}\ and\ \bibinfo {author} {\bibfnamefont {S.}~\bibnamefont
			{Shankaranarayanan}},\ }\href {\doibase 10.1088/1475-7516/2016/10/008}
	{\bibfield  {journal} {\bibinfo  {journal} {JCAP}\ }\textbf {\bibinfo
			{volume} {1610}},\ \bibinfo {pages} {008} (\bibinfo {year}
		{2016}{\natexlab{b}})},\ \Eprint {http://arxiv.org/abs/1606.05747}
	{arXiv:1606.05747 [gr-qc]} \BibitemShut {NoStop}%
\end{thebibliography}
\end{document}